\documentclass[aps, preprint, groupedaddress, floatfix]{revtex4}
\usepackage{epsfig}
\usepackage{graphicx}
\usepackage{amsmath}
\usepackage{amssymb}

\newcommand{\be}{\begin{equation}}
\newcommand{\ee}{\end{equation}}
\newcommand{\ba}{\begin{array}}
\newcommand{\ea}{\end{array}}
\newcommand{\bqa}{\begin{eqnarray}}
\newcommand{\eqa}{\end{eqnarray}}

\newcommand{\etal}{{\it et al.}\ }
\newcommand{\poo}{$p_{\rm O_2}$}
\newcommand{\ie}{{\it i.e.}}
\newcommand{\eg}{{\it e.g.}}
\newcommand{\sto}{SrTiO$_3$\ }
\newcommand{\lao}{LaAlO$_3$\ }
\newcommand{\stolao}{SrTiO$_3$/LaAlO$_3$\ }
\newcommand{\epslao}{\epsilon_{\rm LAO}}

\begin{document}

\title{Electronic and magnetic properties of SrTiO$_3$/LaAlO$_3$ interfaces 
from first principles}

\author{Hanghui Chen$^{1,3}$, Alexie M. Kolpak$^{2,3}$ and
Sohrab Ismail-Beigi$^{1,2,3}$}

\affiliation{$^{1}$ Department of Physics, Yale University, New Haven, CT
06570, USA\\
$^{2}$ Department of Applied Physics, Yale University, New Haven, CT
06570-8284, USA\\
$^{3}$ Center for Research on Interface Structures
and Phenomena (CRISP), Yale University, New Haven, CT 06570-8284, USA}

\date{\today}

\begin{abstract}

A number of intriguing properties emerge upon the formation of the
epitaxial interface between the insulating oxides \lao and
SrTiO$_3$. These properties, which include a quasi
two-dimensional conducting electron gas, low temperature
superconductivity, and magnetism, are not present in the bulk materials,
generating a great deal of interest in the fundamental 
physics of their origins.  While it is generally accepted that the
novel behavior arises as a result of a combination of electronic and
atomic reconstructions and growth-induced defects, the complex
interplay between these effects remains unclear.  In this report, 
we review the progress that has been made towards unraveling the 
complete picture of the \stolao interface, focusing 
primarily on present \textit{ab initio} 
theoretical work and its relation to the experimental
data.  In the process, we highlight some key unresolved issues and
discuss how they might be addressed by future experimental and
theoretical studies.

\end{abstract}

\maketitle

\section{Introduction}

With the advance of techniques to control thin film growth on the
atomic scale, the study of epitaxial oxide heterostructures is a
rapidly developing area of materials science
\cite{Dagotto-Science-2007, Hwang-Science-2006, Reiner-Science-2009}. 
Due to the ability to produce a well-defined, single-terminated surface
\cite{Kawasaki-Science-1994, Koster-APL-1998}, oxide interfaces that
are nearly atomically sharp can now be fabricated.  In many cases, the
properties of these interfaces turn out to be much richer than those
of their bulk constituents \cite{Hwang-Nature-2002, Hwang-Nature-2004,
  Hotta-PRL-2007, Kala-2007, Bousquet-Nature-2007, Taki-PRL-2009}.
Not only are these new interface phases of fundamental physical
interest, they are also promising candidates for novel devices and 
technology.

One of the most interesting epitaxial oxide heterostructures to date
is the (001) interface between \lao (LAO) and \sto (STO) 
\cite{Hwang-Nature-2004}.  
Although both materials
are conventional band insulators in bulk form, among the intriguing
phenomena observed at the \stolao interface is the presence of a
high-mobility quasi two-dimensional electron gas
\cite{Hwang-Nature-2004, Hwang-NatMat-2006}, which emerges when 
the \lao film thickness
reaches a critical value of 1.6~nm (4 \lao unit cells)
\cite{Mannhart-Science-2006, Bell-APL-2009}.
This thickness dependence enables external control of the conductivity
of the heterostructure. Reversible control of the metal-insulator
transition has been demonstrated via an applied electric field
\cite{Mannhart-Science-2006, Triscone-Nature-2008}.  In addition to
these phenomena, at low temperature ($\simeq 200$~mK), the \stolao 
interface becomes superconducting
\cite{Mannhart-Science-2007, Reyren-APL-2009, Schneider-PRB-2009}.  
Furthermore, though bulk \sto and \lao
are both nonmagnetic, experiments suggest that the interface may exhibit 
some type of magnetic ordering \cite{Brinkman-NatMat-2007}.
In addition to fundamental scientific interests, 
the LaAlO$_3$/SrTiO$_3$ interface is also promising for the
development of novel applications in nanoscale oxide electronics. For
example, devices have recently been fabricated \cite{Cen-Science-2009}
that exploit the fact that the interface conductivity can be ``written''
and ``erased'' locally with the aid of an atomic force microscope
(AFM) tip \cite{Cen-NatMat-2008, Blank-NatMat-2008}.

Despite extensive efforts in both theory \cite{Gemming-Mater-2006,
  Pickett-PRB-2006, Freeman-PRB-2006, Albina-PRB-2007,
  Pickett-PRB-2008, Demkov-PRB-2008, Kelly-EPL-2008,
  Satpathy-PRL-2008, Tsymbal-PRL-2009, Chen-PRB-2009,
  Pickett-PRL-2009, Son-PRB-2009, Schuster-EPL-2008,
  Schuster-EPL-2009, Schuster-ChemPhysLett-2009} and experiment
\cite{Willmott-PRL-2007, Vonk-PRB-2007, Kala-PRB-2007,
  Herranz-PRL-2007, Siemons-PRL-2007, Yoshi-PRL-2008,
  Maurice-EPL-2008, Sing-PRL-2009, Salluzzo-PRL-2009, Thiel-PRL-2009},
the origins of these interface properties are not yet completely
resolved. In part, this is due to the fact that more than one mechanism 
may play a role in determining
the interesting behaviors.  Also, which mechanism dominates the behavior
in a given sample appears to be sensitive to the conditions under
which the sample is grown and/or how it is processed prior to the
experimental measurements \cite{Eckstein-NatMat-2007}.  As a result,
direct comparison between experiments can be complicated.
Nevertheless, progress is being made towards understanding the basis
of the novel interface phenomena, in particular, the origin of the 
conductivity.

To date, three mechanisms have been proposed to account for the
emergence of conductivity at the \stolao interface.  The
first mechanism is an intrinsic electronic reconstruction due to the
polar discontinuity at the interface \cite{Hwang-Nature-2004,
  Hwang-NatMat-2006}.  The driving force behind this mechanism can be
understood in the following simple picture. Perovskite oxides have a
generic $ABO_3$ structure, where $A$ and $B$ are metal cations.  Along
the (001) direction, this structure consists of alternating 
$AO$ and $BO_2$ planes. In the simple ionic limit, \sto 
consists of charge neutral layers, (SrO)$^{0}$ and (TiO$_2$)$^{0}$,
while \lao is composed of alternating charged layers, (LaO)$^{+}$
and (AlO$_2$)$^{-}$.  The deposition of stochiometric \lao onto the
\sto (001) substrate is thus equivalent to the formation of a
chain of capacitors in series.  The potential across each capacitor is
additive and, mathematically, diverges with the number of capacitors
(\ie, the thickness of the \lao film).  To offset the
diverging potential, an electronic reconstruction, known as the
``polar catastrophe'', is expected to occur, whereby half an electron
(hole) is transferred to the $n$-type TiO$_2$/LaO ($p$-type SrO/AlO$_2$) 
interface so that the
internal electric field through the \lao film is completely
compensated.  As a consequence of this charge transfer, the interface
becomes doped, leading to the observed conductivity.

The polar catastrophe mechanism is believed to be applicable to the 
\stolao interface because of the multivalent nature of the Ti
cations, which can exist as either Ti$^{3+}$ or Ti$^{4+}$.
Therefore, it is possible that this type of electronic reconstruction
has a lower energy barrier and/or ground state energy than the typical 
atomic reconstructions that occur for more traditional
semiconductor polar/non-polar interfaces \cite{Baraff-PRL-1977, 
Harrison-PRB-1978}.  

However, soon after the polar catastrophe 
mechanism was proposed, several groups \cite{Kala-PRB-2007,
  Herranz-PRL-2007, Siemons-PRL-2007} observed that the electronic
properties of the interface are significantly affected by the presence
of oxygen vacancies, a significant number of which are generated 
in the \sto substrate during the \lao deposition. 
These results suggest that oxygen
vacancies can be responsible for the high sheet
carrier densities and mobilities measured in experiments.  Oxygen
vacancies are also proposed \cite{Cen-NatMat-2008} to account for the
observed insulating-to-metallic transition, via a mechanism involving
the creation and annhilation of oxygen vacancies on the \lao
surface. While the concentration of oxygen vacancies is
suppressed by performing the \lao film deposition under
high oxygen partial pressures and/or post-annealing in oxygen,
some residual conductance is still observed 
\cite{Mannhart-Science-2006, Siemons-PRL-2007}, suggesting that
 oxygen vacancies may not be the sole
source of the interface electron gas.

A third possible mechanism, based on the observed intermixing of
cations across the \stolao interface, has also been
suggested \cite{Hwang-NatMat-2006, Willmott-PRL-2007}.  The driving
force for cation mixing is assumed to be the reduction of the dipole
energy at the interface.  As such mixing results in the formation of
La$_{1-x}$Sr$_x$TiO$_3$ \cite{Willmott-PRL-2007}, which is metallic
 when $0.05 < x < 0.95$ 
\cite{Fuji-PRB-1992, Higu-PRB-2003}, 
it could also lead to interface conductivity.

Though each of the three mechanisms can explain particular
experimental results, there is as yet no unified picture to describe
the entire phase diagram of the \stolao interface and the relations
among the different phases. It is not clear whether the
whole phase diagram has yet been identified, or whether some or all of
these three mechanisms are adequate to account for the
experimental observations.

The primary goals of this report are to review the progress made
towards understanding the \stolao interface using
theoretical first principles calculations and to provide a theoretical
perspective on the existing experimental data.
We start by reviewing the experimental data in Section \ref{Experiment},
highlighting different results in the spirit
of stimulating further studies. Section \ref{Theory} reviews 
the state-of-the-art first-principles
calculations on the \stolao interface and discusses their
successes and failures in describing the experimental observations. In Section
\ref{puzzles}, we pose what we think are the important unanswered questions
about the \stolao interface and in many cases try to suggest additional
experiments and/or theoretical computations.  
Our conclusions are in Section \ref{conclusion}. 
The investigation of \stolao interfaces is a
rapidly growing field. Due to space limitations and the rapid publication 
of new results, we apologize to
those whose work is not included here.

\section{Experiment}
\label{Experiment}

For (001) oriented growth of perovskites, two types 
of \stolao interfaces are possible, depending on 
the sequence arrangement.  The first has the atomic 
planes TiO$_2$/LaO at the interface and is referred to 
as the $n$-type interface; the second is formed by adjacent SrO/AlO$_2$ 
planes and is called the $p$-type interface. Experimentally, it is the 
$n$-type interface that shows the interesting and novel 
properties mentioned in the introduction, while the $p$-type interface 
is found to be insulating.  Below, we briefly review the 
experimental results on the $n$-type and $p$-type interfaces. 
For a more complete experimental review, 
especially concerning the technical details, please refer 
to Refs. \cite{Huijben-AM-2009, Mannhart-MRS-2008}.

\subsection{The $n$-type interface}

\begin{figure*}[t!]
\includegraphics[angle=-90, width=8cm]{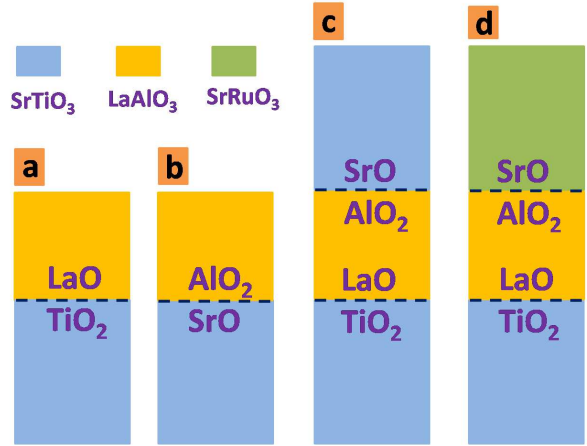}
\caption{\label{fig:expsetup} Schematics of various types of \stolao 
interfaces. The light blue denotes SrTiO$_3$, the orange denotes LaAlO$_3$, 
and the light green denotes SrRuO$_3$. \textbf{a)} A bare $n$-type interface 
(TiO$_2$/LaO) with AlO$_2$ terminated surface. \textbf{b)} A bare $p$-type 
interface (SrO/AlO$_2$) with LaO terminated surface. \textbf{c)} An $np$-type 
interface with \sto as the capping layer. \textbf{d)} A buried $n$-type 
interface with SrRuO$_3$ as the capping layer.}
\end{figure*}

The simplest $n$-type interface (which we call 
the ``bare $n$-type interface'') has a TiO$_2$-terminated 
\sto substrate on which a \lao film is grown, creating a 
TiO$_2$/LaO interface and an exposed AlO$_2$-terminated \lao 
surface \cite{Hwang-Nature-2004, Mannhart-Science-2006} 
(see Fig. \ref{fig:expsetup}a). 
In addition, there are other variants. 
Huijben \etal \cite{Huijben-NatMat-2006}
grew capping layers of \sto on the \lao to form a 
system with both $n$-type and $p$-type interfaces ($np$-type)
(see Fig. \ref{fig:expsetup}c), while Wong \etal \cite{wong-2008} grew 
SrRuO$_3$ electrodes as the 
capping layers on the \lao to study the electric transport 
perpendicular to the interface (see Fig. \ref{fig:expsetup}d).  
These geometries 
create  buried $n$-type interfaces. We briefly review and 
compare the experiments on these $n$-type interface systems.

\subsubsection{Transport} 
\label{Transport}

The transport properties of the $n$-type
\stolao interface depend on the oxygen partial
pressure, \poo, at which the interface is grown, as well as
whether and what type of additional annealing steps are
performed prior to transport measurements.  The existing 
experimental measurements of the sheet resistance $R_s$, 
the sheet carrier density $n_s$, and the Hall mobility 
$\mu_H$, of samples grown at low oxygen partial pressure 
($\sim10^{-6}$ mbar) without post-oxidization appear to have 
converged to the following values:
$R_s \sim 10^{-2}~\Omega$, $n_s \sim 10^{16}-10^{17}$ cm$^{-2}$ and 
$\mu_H\sim 10^4$ cm$^{2}$ V$^{-1}$s$^{-1}$ (see Table~\ref{tab:transport}).

Experiments measuring the transport properties of samples
grown at low oxygen partial pressure ($10^{-5}-10^{-6}$ mbar) and 
then annealed in O$_2$ reveal
a range of results. Ohtomo 
\etal \cite{Hwang-Nature-2004} do not find any significant change in
the transport properties before and after annealing,
whereas Kalabukhov \etal \cite{Kala-PRB-2007} grew
samples under similar conditions, but observe an 
increase in $R_s$ and decreases 
in $n_s$ and $\mu_H$ (see Table~\ref{tab:transport}) after
annealing. 
Basletic \etal \cite{Basletic-NatMat-2008} find similar 
results to those observed by Kalabukhov \etal \cite{Kala-PRB-2007} 
for pre- and post-annealed samples.  

\begin{table*}
\caption{Values of sheet resistance $R_{s}$, sheet carrier density $n_{s}$, 
and Hall mobility $\mu_{H}$ measured from transport
 in different studies. The data 
in this table are all taken at low temperature ($T\sim 5$ K).}
\begin{center}
\begin{tabular*}{1.0\textwidth}{@{\extracolsep{\fill}}cccccc}
\hline\hline
growth $p_{\rm O_2}$(mbar) &annealing conditions &$R_{s}$($\Omega$/$\square$) &$n_{s}$ (cm$^{-2}$) &$\mu_{H}$ (cm$^2$ V$^{-1}$s$^{-1}$) &Ref.\\
\hline\hline
$10^{-6}$ & not annealed                 &$  10^{-2}$        & $  10^{17}$       &$  10^4$ &\cite{Hwang-Nature-2004}\\
$10^{-6}$ & annealed in 1 atm of O$_2$   &$  10^{-2}$        & $ 2\times10^{16}$ &$  10^4$ &\cite{Hwang-Nature-2004}\\
         & at 400$^{\circ}$C for 2 hours &                     &                 &             &                        \\
$10^{-6}$ & not annealed                  &$  10^{-2}$       &$ 5\times10^{16}$ &$  10^4$ &\cite{Kala-PRB-2007}    \\
$10^{-6}$ & annealed at 500 mbar          & $  10^3$         &$ 10^{13}$     &$  10^3$ &\cite{Kala-PRB-2007}    \\
        & during cooling                &                      &                 &            &                        \\
$10^{-6}$ & annealed at 300 mbar         & $ 10^2$ &$ 3\times 10^{13}$ &$  7\times 10^{2}$&\cite{Basletic-NatMat-2008}\\
         & during cooling               & & & & \\
$10^{-6}$ & not annealed                 & $  3\times 10^{-3}$&                 &$  10^{4}$&\cite{Herranz-PRL-2007}\\
$10^{-6}$ & not annealed                 &                   &$ 2\times 10^{16}$&$  10^{4}$&\cite{Siemons-PRL-2007}\\     

\hline
$10^{-5}$ & not annealed                &$  10^{2} $   & $ 10^{16}$    & $  10^{0}$ &\cite{Hwang-Nature-2004} \\
$10^{-5}$ & not annealed                &$  10^{-3}$   & $ 5\times 10^{17}$& $  2\times10^{3}$&\cite{Basletic-NatMat-2008}\\
$2\times10^{-5}$ & cooled at high oxidation &             & $  2\times 10^{13}$ & $  3\times 10^2$ &\cite{Siemons-PRL-2007}\\
\hline
$10^{-4}$ & not annealed                &$  10^4$            &$ 10^{14}$     &$  10^{1}$  &\cite{Hwang-Nature-2004} \\
$10^{-4}$ & not annealed                &$  10^2$            &$ 10^{13}$     &$  10^3$    &\cite{Kala-PRB-2007}     \\
$10^{-4}$ & annealed at 500 mbar        &$  10^2$            &$ 10^{13}$     &$  10^{3}$  &\cite{Kala-PRB-2007}     \\
         & during cooling              &                       &                 &               &                         \\
$10^{-4}$ & not annealed                &$  3\times10^{-3}$   &                 &$  10^{1}$  &\cite{Herranz-PRL-2007} \\
\hline
$5\times 10^{-2}$ & not annealed        & insulating            &insulating       &insulating     &\cite{Kala-PRB-2007}       \\
$10^{-6}-10^{-3}$ & annealed at 300 mbar & insulating           &insulating       &insulating      &\cite{Herranz-PRL-2007}   \\
                & during cooling       &                      &                 &                &                          \\
\hline\hline
\end{tabular*}
\end{center}
\label{tab:transport}
\end{table*}

There appears to be general agreement that, for samples grown 
at high oxygen partial pressures ($\sim 10^{-4}$
mbar), the sheet carrier density is in the range 
$10^{13}-10^{14}$ cm$^{-2}$. Some discrepancies arise
in mobility measurements of the same type of samples: at low 
temperatures ($\sim 5$ K), 
$R_s \sim 10^4~\Omega$ and $\mu_H \sim 10$ cm$^2$ V$^{-1}$s$^{-1}$
are obtained by Ohtomo \etal~\cite{Hwang-Nature-2004} and 
Herranz \etal \cite{Herranz-PRL-2007}, while Kalabukhov 
\etal \cite{Kala-PRB-2007} and Thiel \etal \cite{Mannhart-Science-2007} 
report $R_s$ and $\mu_H$ values two orders of magnitude smaller 
and larger, respectively (see Table \ref{tab:transport}).

It is noteworthy that the polar catastrophe mechanism 
\cite{Hwang-Nature-2004, Hwang-NatMat-2006} 
predicts that 0.5 electrons per two-dimensional interface 
unit cell ($0.5e/a_{\textrm{STO}}^2$, where $a_{\textrm{STO}}$ is the 
lattice constant of SrTiO$_3$) at the
$n$-type interface would completely cancel the internal electric field
through the LaAlO$_3$ film. That is equivalent to a sheet carrier density of
$3.3\times10^{14}$ cm$^{-2}$, implying that the polar catastrophe should be
the dominant mechanism determining the sheet carrier density in 
samples grown at high oxygen partial pressures ($\sim 10^{-4}$ 
mbar), while other mechanisms, such as
doping via oxygen vacancies, may be responsible for the
higher sheet carrier densities of interfaces grown at lower oxygen partial 
pressures ($\sim 10^{-6}$ mbar).  However, agreement between the 
predicted and measured sheet carrier densities in samples grown at high oxygen
partial pressure is not sufficient to rule out the possibility
that other mechanisms, such as oxygen vacancies on the \lao
surface \cite{Cen-NatMat-2008}, also play a role.  These other 
possibilities  may shed some light on the unexplained 
discrepancies in the measured
mobilities in these samples.  

We note that samples grown at
very high oxygen partial pressures ($>10^{-2}$ mbar) 
\cite{Kala-PRB-2007} or samples annealed at high oxygen 
partial pressure \cite{Herranz-PRL-2007} have been observed
to be completely insulating. If post-annealing in oxygen
fills oxygen vacancies and repairs the oxygen 
off-stoichiometry, then one would expect that with 
increasing oxygen partial pressure, the sheet carrier 
density would asymptote to $\sim 10^{14}$ cm$^{-2}$, 
the value predicted for polar 
\lao and an ideal $n$-type interface as per 
the polar catastrophe mechanism.  
We will further discuss this situation 
in Section \ref{insulating n-type}.

Another factor affecting transport is the 
thickness of the \lao films.  Samples grown under 
relatively high oxygen partial
pressure ($\sim 10^{-4}$ mbar) are not conducting unless the \lao 
film thickness exceeds a critical value, referred to 
as the ``critical separation''.  The two studies 
investigating the thickness dependence of sheet 
carrier density report critical separations of 4 and 6 unit cells
\cite{Mannhart-Science-2006,Siemons-PRL-2007}, respectively. 

\subsubsection{Spatial extent and orbital character of the electron gas} 

The oxygen partial pressure during
the growth process affects not only the transport properties but also
the spatial extent of the electron gas. 
Two different distributions of the
electron gas have been observed.  For samples grown at low
oxygen partial pressure ($\sim$10$^{-6}$ mbar) and not 
subjected to further annealing, the electron gas extends 
over tens to hundreds of $\mu$m into the
\sto substrate and thus has 3D character. These electrons most 
likely originate from extrinsic dopants (oxygen vacancies) 
throughout the \sto substrate. On the other hand, samples 
grown and/or annealed at a higher oxygen partial pressure 
($\sim 10^{-4}$ mbar) have conducting electron gases that 
are quasi-2D, confined within a few (to at most 
tens of) nanometers of the interface. These results on the thickness 
of the electron gas are based on a number of methods, 
including solution of the Poisson equation for the interface 
system \cite{Siemons-PRL-2007}, observation of magnetoresistance 
and Shubnikov-de Hass oscillations at the interface 
\cite{Herranz-PRL-2007}, direct imaging using contact-tip atomic 
force microscopy \cite{Basletic-NatMat-2008}, hard x-ray
 photoelectron spectroscopy \cite{Sing-PRL-2009}, and modeling 
of the superconducting properties of the interface electron gas 
\cite{Mannhart-Science-2007}.

In addition, experiments employing x-ray absorption 
spectroscopy \cite{Salluzzo-PRL-2009} have been able 
to furnish information on the orbital character of the 
electron gas at the $n$-type interface. Specifically, 
the degeneracy of $t_{2g}$ states (originating from the 
Ti atom in SrTiO$_3$) is removed, and the lowest states for conducting 
electrons are of Ti $d_{xy}$ character.

\subsubsection{Thickness dependence of sheet carrier density} 

\begin{figure*}[t!]
\includegraphics[angle=-90, width=11.6cm]{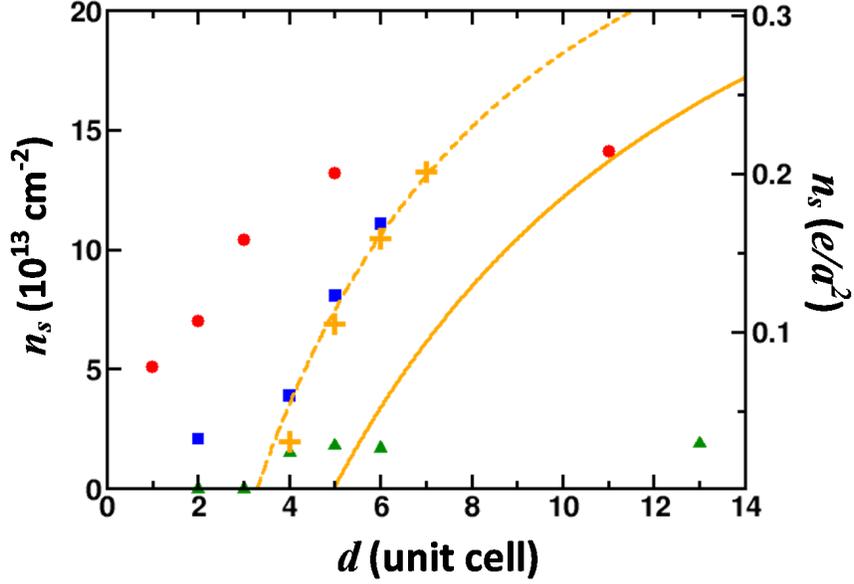}
\caption{\label{fig:sdensity} A comparison of sheet carrier density at
  the $n$-type interface from different published experiments as well
  as those from first principles theory. The red circles are data from
  Ref. \cite{Huijben-NatMat-2006}, the blue squares are data from
  Ref. \cite{Sing-PRL-2009} and the green triangles are data from
  Ref. \cite{Mannhart-Science-2006}. The data from
  Ref. \cite{Huijben-NatMat-2006} and
  Ref. \cite{Mannhart-Science-2006} are extracted from transport
  measurements. The data from Ref. \cite{Sing-PRL-2009} are obtained
  using hard x-ray photoelectron spectroscopy.  Orange pluses are
  values directly obtained from first principles DFT-LDA calculations.
  The dashed and solid curves are fits to the first principles data
  where the \lao is treated as a continuous medium with field
  dependent dielectric constant (see Sec. \ref{sec:thycritsep}.)  The
  dashed curve is a model using the bulk DFT-LDA band gaps for \sto
  and LaAlO$_3$ whereas the solid curve is corrected for the
  underestimation of the bulk DFT-LDA band gaps.}
\end{figure*}

While it is generally agreed upon that there is a minimum required
thickness for the \lao film before mobile carriers and thus
conductivity are observed at the $n$-type interface, the details 
of this behavior raise issues yet to be resolved. 
As with the transport properties,
in general, the thickness dependence of $n_{\rm s}$ is sensitive to
the sample growth conditions. A comparison of the sheet carrier density
determined from different available experiments is shown in
Fig. \ref{fig:sdensity}, along with theoretical predictions from our
first principles calculations.  (Similar theoretical curves are
reported by Son \etal \cite{Son-PRB-2009}.)

Thiel \etal
\cite{Mannhart-Science-2006} show that after annealing, the
sheet carrier density for samples with fewer than 4 unit cells (u.c.)
 of \lao is below detectable levels. Above 4 u.c., they observe 
a step-like jump in $n_s$, which then remains in the range of
$1\times 10^{13} -2\times 10^{13}$ cm$^{-2}$ and shows no further  
discernible thickness dependence.  The discontinuous jump in sheet 
density agrees with the polar catastrophe model of the 
metal-insulator transition, but the value of the sheet density 
is an order of magnitude smaller than the theoretical value of 
$0.5e/a_{\textrm{STO}}^2$, or $3.3\times10^{14}$ cm$^{-2}$. 

Huijben \etal \cite{Huijben-NatMat-2006} grew $np$-type 
interface samples with \poo\ of $10^{-5}$ mbar without annealing 
and observe that
the samples are conducting for all \lao thicknesses studied: 
$n_s$ starts out small but finite 
for 1 u.c. of \lao and increases with \lao thickness until it 
saturates at $1.4\times 10^{14}$ cm$^{-2}$ for $\ge$6 u.c. of 
LaAlO$_3$.  
As Fig. \ref{fig:sdensity} shows, 
here the saturation sheet density is closer to the 
theoretically expected value of $0.5 e/a^2_{\textrm{STO}}$, 
but the carrier density is 
nonzero and significant for all \lao thicknesses, which is 
at variance with a metal-insulator transition scenario.

The above results are obtained from transport measurements.
Sing \etal \cite{Sing-PRL-2009} employ hard x-ray photoelectron 
spectroscopy to study samples grown under the same conditions as 
Thiel \etal \cite{Mannhart-Science-2006} but find that the carrier 
density is already non-zero for \lao films 2 u.c. thick.
In contrast to
Ref.~\cite{Mannhart-Science-2006} but in qualitative 
agreement with
Ref.~\cite{Huijben-NatMat-2006} (despite the differing growth
 conditions), they observe that $n_{\rm s}$ increases with the 
\lao thickness. The authors propose that the finite sheet
 carrier density of samples with 2 u.c. of \lao is immobile and 
does not contribute to transport. This raises the interesting 
and important question of whether some or all electrons at the 
$n$-type interface participate in transport, which we further 
discuss in Section \ref{Anderson}.

\subsubsection{Cation disorder and intermixing}

Despite the high level of epitaxy available with modern growth
techniques, some research groups have observed that significant cation
mixing occurs at the $n$-type interface.  Nakagawa 
\etal \cite{Hwang-NatMat-2006}, using atomic-resolution electron 
energy loss spectroscopy (EELS), find that the $n$-type
 interface is significantly rougher than the $p$-type interface. 
They suggest that the finite width of the electron gas 
(as dictated by basic quantum mechanics) increases the 
electric dipole energy, which in turn is reduced by 
exchanging Sr with La across the interface.  Willmott 
\etal \cite{Willmott-PRL-2007} analyze the interface structure using
coherent Bragg rod analysis (COBRA) and also observe a graded
intermixing over 3 unit cells. In addition, they note that La-Sr
mixing is evident for $\sim$6~\AA~further away from the interface than
Al-Ti mixing. Since the ionic radii of La and Sr are more than twice 
as large as those of Al and Ti, the observation of the intermixing 
is interpreted as a natural tendency of the system, rather than an 
effect due to kinetic disruption during growth.  Overall, the basic 
mechanisms that are the driving forces for cation disorder require
 explanation and exploration, as does the question of what procedures 
can effectively reduce such intermixing to create ideal interfaces.

\subsubsection{Magnetic properties} 

Experimental studies of the magnetic properties of the $n$-type 
interface are an active and growing area.  To date, some intriguing
 discoveries have been reported by various research teams based on 
using magnetoresistance measurements as an indirect probe of magnetic
 behavior and ordering. 
Measurements by Brinkman \etal \cite{Brinkman-NatMat-2007} 
provide the first experimental
evidence of magnetic behavior.  Since then, additional studies 
\cite{vanzalk-2008, shalom-2009} suggest that some type of magnetic
ordering may occur at low temperatures.  However, there are still 
questions about whether the system exhibits hysteretic behavior, 
the sign of the magnetoresistance,
and whether the magnetoresistance is 
isotropic or anisotropic.

Brinkman \etal \cite{Brinkman-NatMat-2007} studied samples grown 
under a relatively high \poo\ $\sim10^{-3}$ mbar, with
sheet resistances $>10^{4}~\Omega$. They observe isotropic 
magnetoresistance, which they ascribe to spin
scattering off localized magnetic moments at the interface,
 an interpretation bolstered by their observation of a minimum 
in the sheet resistance $R_s(T)$ with lowering temperature.  
This minimum is reminiscent of the Kondo effect, which stems 
from the interplay between localized magnetic moments and 
itinerant charges.  Furthermore, they find that the magnetoresistance 
at 300 mK is hysteretic, implying the
presence of ferromagnetic domains.  However, the $R_s(T)$ 
minimum is not present for samples grown at lower 
\poo\ ($< 10^{-5}$ mbar). 
In contrast, Reyren \etal \cite{Mannhart-Science-2007} grew 
a sample at a low oxygen partial pressure of $6 \times
10^{-5}$~mbar, which is then annealed under conditions
 that (presumably) yield well oxidized samples.  They find 
positive magnetoresistance, no minimum in $R_s(T)$, 
and no hysteresis in magnetoresistance. The two major 
differences between the experiments of Refs.~\cite{Brinkman-NatMat-2007}
and \cite{Mannhart-Science-2007} are the growth \poo\ 
($10^{-3}$ mbar
versus $10^{-5}$ mbar) and the thickness of the 
\lao film
(26 u.c versus 8 u.c.). The former factor is known to 
affect the sheet resistance and mobility, while 
the latter modifies the sheet carrier density. 

The \stolao samples investigated by van Zalk \etal 
\cite{vanzalk-2008} are grown under similar conditions to those
used by Brinkman \etal~\cite{Brinkman-NatMat-2007}.  This 
study reports hysteresis below 300 mK and magnetoresistance 
oscillations versus external field intensity $B$ at 50 mK.  
Surprisingly, the oscillations are periodic in $\sqrt{B}$  
instead of the standard Shubnikov-de Haas $1/B$ periodicity. 
The authors suggest that this stems from
the commensurability condition of edge states. 
They also speculate 
that the magnetoresistance oscillation is related to
ferromagnetic ordering, and that the quantum Hall effect might be present.

Shalom {\it et al.} \cite{shalom-2009} studied samples 
grown at a lower \poo\ of $10^{-4}-10^{-5}$~mbar.  In agreement 
with other observations \cite{Kala-PRB-2007, Mannhart-Science-2006, 
Mannhart-Science-2007}, the sheet carrier density 
is $\sim$$10^{13}$ cm$^{-2}$ and the system becomes
 superconducting at 130 mK. In
contrast to Refs.~\cite{Brinkman-NatMat-2007} and 
\cite{vanzalk-2008}, however,
the low temperature magnetoresistances are highly 
anisotropic and have different signs for fields parallel 
or perpendicular to the current.
They find no hysteresis down to 130 mK, at 
which point their samples become superconducting.  Based on
these observations, they suggest an anti-ferromagnetic ordering
at the interface below a N\'{e}el temperature of $\sim$35~K.

\subsection{The $p$-type interface}

Experimentally, the $p$-type interface (AlO$_2$/SrO) is less
explored than the $n$-type
interface. 
Nakagawa \etal \cite{Hwang-NatMat-2006} suggest 
that the insulating properties of the $p$-type interface 
arise from the presence of oxygen vacancies.
 Unlike the $n$-type inteface, at which the Ti 
can exist in multiple valence states, the $p$-type 
interface has no multi-valent element (\eg, O$^{-}$ 
is very unusual in compounds).  Hence, they argue 
that, in order to compensate the internal polar field 
and potential across the \lao film, an atomic 
reconstruction such as a high density of oxygen vacancies is required.

In a simple ionic picture, an ideal and unreconstructed $p$-type 
interface has 0.5 holes per interface unit cell to compensate the 
polar field in the \lao film.  An oxygen vacancy can potentially 
donate two electrons, and assuming that both electrons are mobile, a 
density of one oxygen vacancy per four two-dimensional (2D) 
interface unit cells can precisely accommodate the holes and 
leave no
carriers at the $p$-type interface.  Using atomic resolution 
EELS and various fitting procedures, Nakagawa \etal 
\cite{Hwang-NatMat-2006} infer the presence of $0.32 \pm 0.06$ 
oxygen vacancies per 2D unit cell in the \sto near the $p$-type interface, 
which is three times higher than their inferred value of 0.1 oxygen 
vacancies per 2D unit cell at the $n$-type interface. 

Although these results support the oxygen vacancy picture, 
they also raise some questions. The first issue is whether
 both electrons on the oxygen 
vacancy are in fact {\it mobile} and hence can be donated 
to the interface. The experimental 
finding of 0.32 oxygen vacancies per 2D unit cell argues that more 
than 0.25 oxygen vacancies per 2D unit cell are needed. There 
is also theoretical evidence to support this incomplete 
donation of electrons \cite{Freeman-PRB-2006}, 
a subject we return to in Section \ref{ov at ptype}.

Regardless of the precise number of oxygen vacancies, 
in the polar catastrophe scenario the holes are created 
by $0.5e/a_{\textrm{STO}}^2$ electrons being transfered away from the $p$-type 
interface to the other side of the \lao film, whether it 
be the \lao surface or another $n$-type interface with 
a capping layer.  Why do these electrons make no 
contribution to the transport?  This question also 
applies to the $n$-type and $np$-type interfaces, which similarly should 
have holes on the \lao surface or at the capping $p$-type 
interface, but experimentally show only electron-like 
carriers in transport.  We discuss this point further in Section 
\ref{surface conductivity} below. 

In summary, though the $n$-type interface has interesting
properties that are promising for potential applications,
understanding the $p$-type interface is equally important from a
theoretical point of view. 

\section{First Principles Theory}
\label{Theory}

In this section, we focus on the theoretical work on the \stolao interface 
system. Most of the theoretical literature is devoted to elucidating 
the origin of the conducting electron gas, reproducing the 
insulating-to-metallic transition, and designing new 
interfaces with more interesting properties. Some successes have
 been 
achieved in the comparison between theory and experiment. 
However, discrepancies exist and some important 
questions are unresolved. Although some interesting results 
have been obtained from model Hamiltonian calculations
\cite{Lee-PRB-2007}, we  
concentrate here on first principle calculations.

\subsection{Theoretical methodologies}

With few exceptions, most first principles studies of \stolao
 interfaces to date have focused on defect-free and idealized 
$n$-type and $p$-type interfaces.  Aside from the desire to 
understand the basic physics and intrinsic properties of high 
quality interfaces, the primary practical reason has been due 
to computational costs: including imperfections such as oxygen 
vacancies, cation intermixtures, or other atomic reconstructions 
requires the use of relatively large simulation cells.
 Theoretical studies reporting on oxygen vacancies include 
Refs.~\cite{Pickett-PRB-2006,Freeman-PRB-2006,Cen-NatMat-2008,Chen-PRB-2009} 
with one work on modeling intermixing \cite{Willmott-PRL-2007}.  
However, understanding the ideal interfaces is both a good 
starting point for gaining theoretical insight and should 
also be relevant to experiments where it is believed that 
high quality interfaces have been achieved.

Most of the reported calculations use density functional 
theory 
(DFT) \cite{Kohn-PR-1965} with the local density approximation (LDA) 
\cite{Kohn-PR-1965} or the generalized 
gradient approximation (GGA) \cite{Perdew-PRL-1996}. 
A subset have employed the LDA+U 
method to include approximately some effects of localized 
electronic correlations that are missing from 
standard LDA or GGA DFT 
calculations \cite{Anisimov-Condense-1997}.  
This has been done typically by including Hubbard $U$ and 
exchange $J$ corrections on the Ti-$d$ manifold, although 
Pentecheva and Pickett 
\cite{Pickett-PRB-2006} apply $U$ on the oxygen $p$ orbitals 
to study the possibility of hole polarons at the $p$-type interface. 
The $U$ values employed and the resulting predictions 
are not unianmous, ranging from 5 eV to 8 eV \cite{Kelly-EPL-2008, Pickett-PRB-2006, Pickett-PRB-2008, Tsymbal-JAP-2008, Demkov-PRB-2008}.
In particular, Zhong and Kelly \cite{Kelly-EPL-2008} 
studied the $n$-type interface properties as 
a function of $U$ ranging from 0 to above 6 eV and 
find a variety of geometric, electronic, and magnetic ground 
states versus $U$. 

\begin{figure*}[t!]
\includegraphics[angle=0, width=15cm]{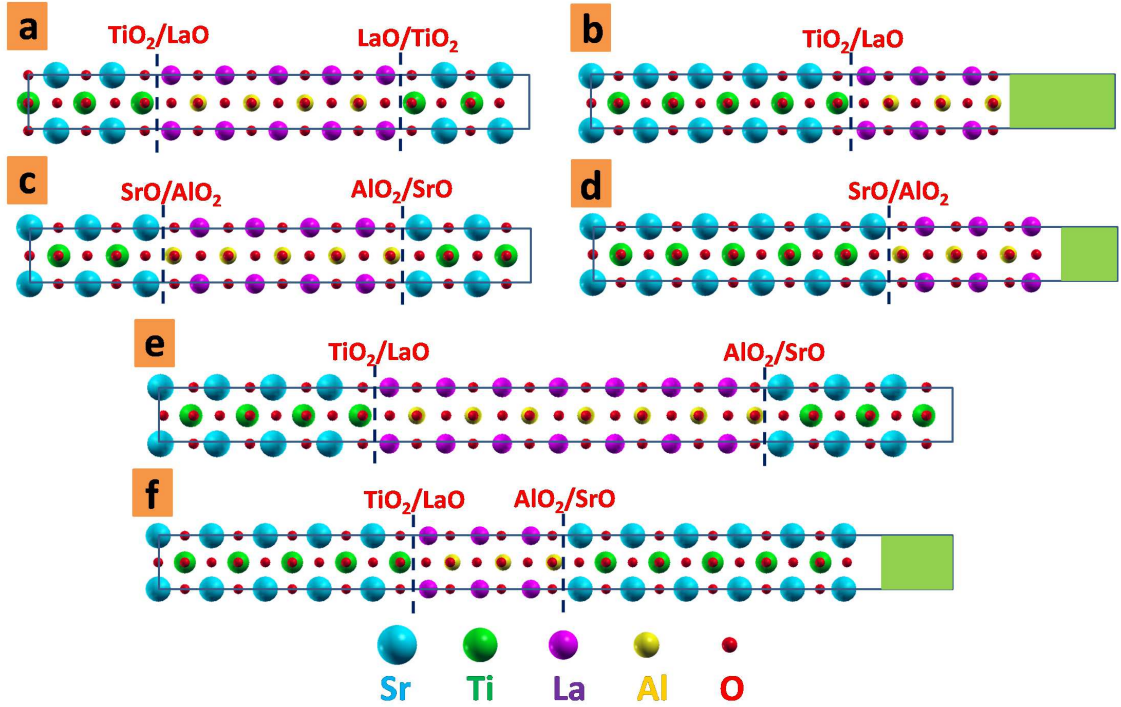}
\caption{\label{fig:config} Schematics of simulation cells for various types
  of interfaces and superlattices for the \stolao system. The shaded
  green regions denote vacuum; their thickness is merely
  schematic and not indicative of actual thickness used in the simulation. 
  For superlattice
  simulations, there is no vacuum in the unit cell. For interface
  simulations, the vacuum is inserted to separate periodic copies and
  introduce surfaces.  \textbf{a)} $n$-type superlattices, \textbf{b)}
  $n$-type interfaces, \textbf{c)} $p$-type superlattices, \textbf{d)}
  $p$-type interfaces, \textbf{e)} $np$-type superlattices and
  \textbf{f)} $np$-type interfaces.}
\end{figure*}

The simulation cells used fall into three distinct 
categories, which provide complementary information on these 
interfaces. The most popular approach has been to employ 
a symmetric superlattice:  a number of unit cells of \sto are 
placed next to a number of unit cells of \lao along (001) and 
periodic boundary conditions are applied.  In this symmetric 
superlattice approach, both the \sto and \lao films 
are nonstoichiometric: when the \sto has an additional TiO$_2$ 
atomic layer and the \lao has an extra LaO layer, two identical 
$n$-type interfaces are formed to make a symmetric system 
(Fig. \ref{fig:config}a); 
extra SrO and AlO$_2$ layers create two $p$-type interfaces 
(Fig. \ref{fig:config}c). The symmetric 
superlattice allows for simulation of a single interface type 
in isolation, and leads to relatively small simulation cells.  
However, due to the imposed symmetry and the lack of stoichiometry, 
the \lao film is nonpolar.  This precludes the study of the polar 
catastrophe and the evolution of the interface properties as a 
function of \lao thickness. 
Due to the nonstoichiometry, one additional electron (for double 
$n$-type) or one additional hole (for double $p$-type) is introduced 
into the symmetric supercell, which means essentially 0.5 carriers 
per 2D interface unit cell are doped at each interface in the 
calculation, \ie, the sheet carrier density is fixed as 
dictated by chemistry. 
Physically, 0.5 carriers per 2D cell 
is the required amount to completely compensate the \lao 
polar field \cite{Hwang-NatMat-2006}, meaning that the 
symmetric supercell approach provides the properties 
of the interfaces in the thick \lao film limit.

Experimental growth procedures generate 
stoichiometric \lao and \sto films that are thin 
(on the order of nanometers) and often have exposed oxide surfaces or 
capping layers.  Therefore, the next generation of computations 
employs simulation cells with stoichiometric films and regions of 
vacuum to break periodic boundary conditions along (001), 
as illustrated in Fig. \ref{fig:config}b, \ref{fig:config}d, and 
\ref{fig:config}f.  
The presence of a vacuum region in the 
simulation cell allows for the polar potential developed through
 the \lao film to drop across the vacuum gap instead of through the 
materials themselves. To simulate the stoichiometric 
interfaces, a slab of \sto is typically used to model the substrate, 
a \lao film is placed on this slab, and, depending on the system 
being modeled, the \lao can terminate at the vacuum or a further 
capping film is added on top of the LaAlO$_3$. In the first case, 
the system has a single buried $n$-type (Fig. \ref{fig:config}b) 
or $p$-type (Fig. \ref{fig:config}d) interface with 
an exposed surface; in the second case, a pair of $n$-type and 
$p$-type interfaces coexist on the two sides of the \lao film.  
We refer to these system as the $n$-type, $p$-type, and $np$-type 
interfaces, respectively.  (We note that all three types have been 
fabricated in experiments.)  These stoichiometric simulations have 
a polar \lao film with an internal field going from the (LaO)$^+$ 
to (AlO$_2$)$^-$ terminating atomic layers.  Therefore, such 
simulations can probe the evolution of the polar catastrophe 
and sheet carrier properties with \lao thickness. Furthermore, 
assuming ideal interfaces and surfaces, these calculations 
represent a faithful reproduction of many of the actual 
experiments. Due to the close proximity of the exposed 
surfaces to the interfaces, the effect of surface defects or 
adsorbates on the interface properties can also be investigated.

Finally, there is another type of geometry employed in 
simulations \cite{Freeman-PRB-2006, Albina-PRB-2007, Gemming-Mater-2006, 
Bristowe-PRB-2009}, which we refer to as 
$np$-type superlattices (Fig. \ref{fig:config}e). 
These superlattices have stoichiometric \sto and \lao 
but do not include vacuum. 
The absence of vacuum makes the electronic structures of 
$np$-type superlattices different from those 
of $np$-type interfaces (Fig. \ref{fig:config}f), as we 
discuss in more detail in Section \ref{sec:thycritsep} with more details.

\subsection{The $n$-type interface}

In this subsection, we review the state-of-art theoretical 
results from first principles simulations of $n$-type and $np$-type 
interfaces, as well as double $n$-type and $np$-type 
superlattices.  

\subsubsection{Critical separation}
\label{sec:thycritsep}
\begin{figure*}[t!]
\includegraphics[angle=0, width=14cm]{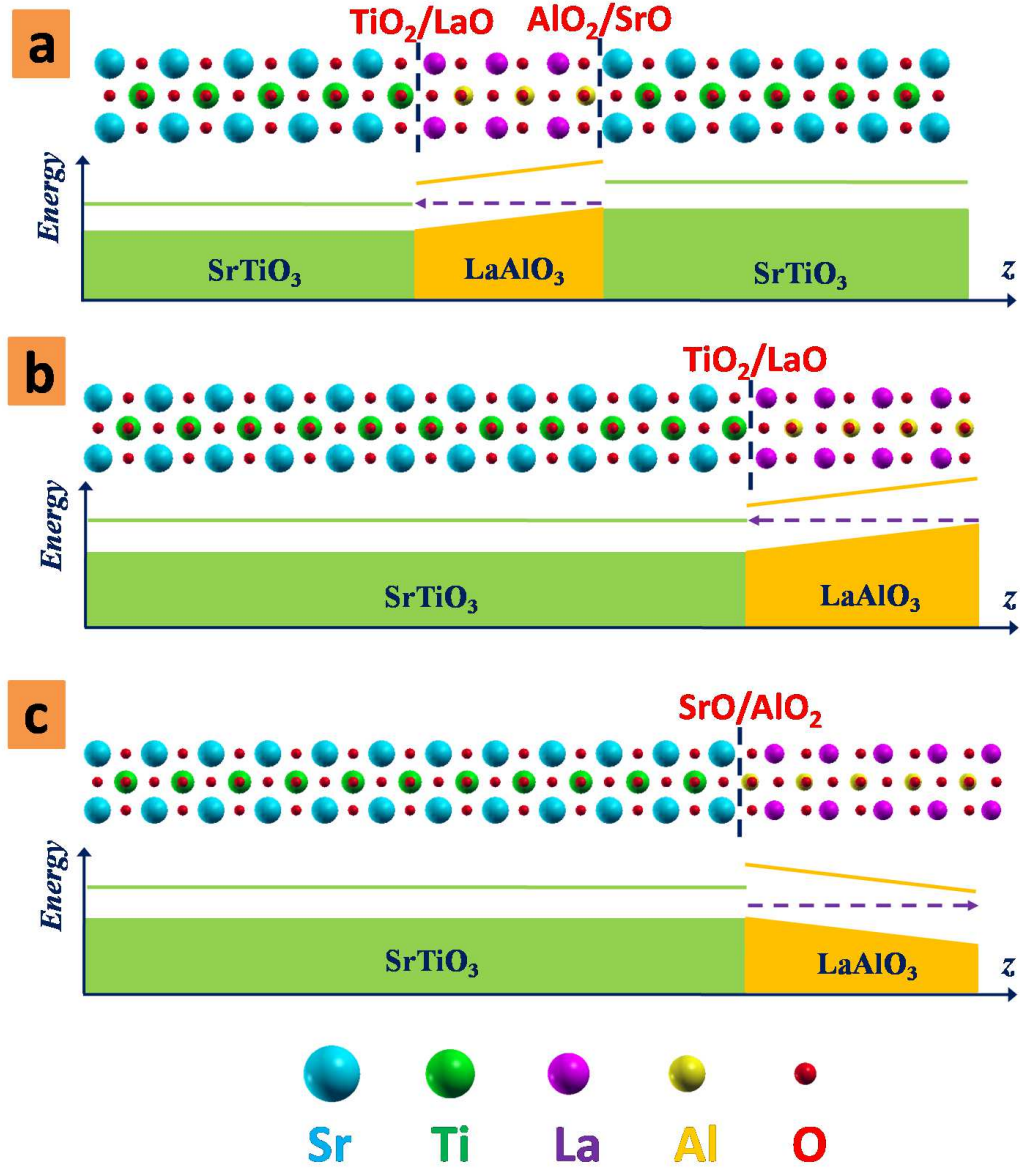}
\caption{\label{fig:ediagram} Schematic of energy diagrams for \textbf{a)} 
$np$-type, \textbf{b}) $n$-type and \textbf{c)} $p$-type interfaces, 
respectively. The shaded parts are filled valence 
bands. The conduction bands are empty, and their edges 
are denoted by the solid lines. The dashed arrow refers to the direction 
of charge transfer.}
\end{figure*}

As discussed above, an insulating-to-metallic transition has been
observed at the \stolao interface for \lao films
thicker than 4 u.c. \cite{Mannhart-Science-2006}.
The polar nature of \lao provides an explanation for this
behavior, \ie\ the polar catastrophe.  This transition 
can be understood by considering the \stolao energy band diagram.
The energy diagram of the $n$-type interface is shown in
Fig. \ref{fig:ediagram}b. Drawn schematically on the left side 
are the valence and
conduction band edges of the \sto substrate, which are composed
of O $p$- and Ti $d$-states, respectively.  To the right are the
\lao band edges, which are composed of O $p$- and La $d$-states, 
respectively.  Due to the polar nature of LaAlO$_3$, there is an 
electric field through the stoichiometric \lao film. Consequently, 
the bands slope up linearly inside the \lao film.  The upwards 
slope in Fig. \ref{fig:ediagram}b indicates that the positively 
charged (LaO)$^+$ 
layer at the $n$-type interface attracts electrons.  As 
the figure illustrates, the
energy gap of the entire $n$-type interface system is 
determined by the difference between conduction band edge 
Ti~$d$-states of the \sto substrate and the valence band 
edge O~$p$-states at the \lao surface.  Each added \lao 
unit cell increases the surface valence band edge 
by approximately a fixed amount, thereby reducing the total 
energy gap of the entire interface system.  The smallest 
thickness of \lao that closes the energy gap is thus the 
critical separation at which the system becomes metallic.  
Fig. \ref{fig:ediagram}a shows that the $np$-type interface behaves 
similarly, 
although in this case the total energy gap is determined 
by the conduction band edge Ti $d$-states of the \sto 
substrate and the valence band edge O $p$-states in the capping \sto layer.

The insulating-to-metallic transition has been
directly confirmed by DFT studies on stoichiometric interfaces such as
those shown in Fig.~\ref{fig:config}b and \ref{fig:config}f 
\cite{Demkov-PRB-2008, Chen-PRB-2009, Pickett-PRL-2009, Son-PRB-2009}.  
By computing local densities 
of states in each
atomic plane, it is straightforward to locate the band edges and to
recreate energy band diagrams of the style shown in Fig. \ref{fig:ediagram}a 
and \ref{fig:ediagram}b.  
These diagrams show precisely the behavior described above.  

While the theoretically predicted and experimentally observed 
values for the critical separation are similar, they are not in 
quantitative agreement. From the discussion above, one can see that 
the value of the critical separation depends on the band gap of 
SrTiO$_3$. As is well known, band gaps are systematically underestimated 
in DFT calculations.  Nevertheless, direct comparison with 
experiment can be achieved by including a manual correction 
to the band gap, which physically amounts to adding 1 or 2 
unit cells of \lao to the theoretical critical separation.  
In previous work~\cite{Chen-PRB-2009}, we employ LDA to 
study the
critical separation of the $n$-type and $np$-type interfaces. 
 We determine that the increase of the potential with each 
added \lao unit cell is $\simeq 0.7$ eV, which corresponds to an average 
electric field of 0.19~V/\AA\ in the \lao film prior 
to the polar catastrophe. The band gap of \sto in our calculations
is 1.9 eV. Taking into account the difference between this and the
experimental \sto band gap of 3.2 eV, we predict
that the experimental critical separation is 6 and 5 \lao u.c. for
the $n$-type and $np$-type interfaces, respectively.  Using GGA with a
similarly underestimated \sto band gap of 2.0~eV, Pentcheva and 
Pickett \cite{Pickett-PRL-2009} predict a critical separation 
of greater than 6 unit cells for the $n$-type interface based 
on a potential shift of 0.4 eV per added \lao unit cell. 
Cen \etal \cite{Cen-NatMat-2008} find similar values as well: using GGA
calculations with a \sto band gap of 1.8 eV, they find the
theoretical critical separation of the $n$-type interface to be 4 u.c.
of \lao and predict that the experimental critical
separation should be larger than this value.  Therefore, 
for ideal $n$-type interfaces and polar \lao films, LDA 
and GGA-based DFT calculations have reached a consensus 
that the critical thickness is larger than 4 unit cells 
and is most likely 6 unit cells for the $n$-type interface.  

Comparison to experiment is complicated by the experimental 
variances discussed in the previous section. The theoretical predictions are 
larger than 
the experimental value of 4 u.c. of Thiel \etal 
\cite{Mannhart-Science-2006} as determined by transport, or even 
2 u.c. from hard x-ray spectroscopy results of Sing \etal 
\cite{Sing-PRL-2009} for $n$-type interfaces with exposed 
\lao surfaces.  For the buried $np$-type interfaces 
studied by Huijben \etal  \cite{Huijben-NatMat-2006}, 
the comparison is better in the sense that the sheet carrier 
density saturates for \lao films 6 u.c. thick.  However, those 
experiments find finite carrier densities
 even for 1 u.c. of LaAlO$_3$, in disagreement with the metal-insulator 
transition scenario.

At the LDA+U level, Lee and Demkov \cite{Demkov-PRB-2008} use
a Hubbard correction on the Ti $d$ manifold with $U$=8.5 eV 
to shift the conduction band edge of \sto and increase its 
band gap to 3.2 eV.  They report a stronger polar field of 
0.24~V/\AA.  They simulate three and five u.c. 
thick \lao films and find that the five unit cell case undergoes an 
metal-insulator transition. Therefore, they report a smaller 
critical separation of 4 or 5 \lao unit cells. This value 
is closer to the experimental one determined from transport measurements 
on the $n$-type 
interface with exposed \lao surfaces of Ref.~\cite{Mannhart-Science-2006} 
but differs from the other two experiments.

\begin{figure*}[t!]
\includegraphics[angle=0, width=10cm]{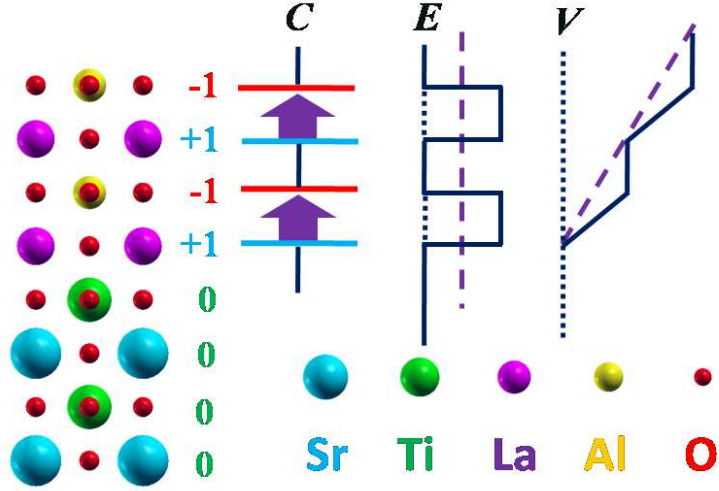}
\caption{\label{fig:average} 
Schematic showing how an \lao film can be approximated as a series of 
parallel-plate capacitors and the resulting distribution of electric field 
and potential. The solid lines refer to the actual electric field 
and potential generated by the capacitors. The dashed lines are the 
averaged electric field and potential over each unit cell.}
\end{figure*}

We note that in all the above simulations, the \lao film itself is
modeled using standard LDA or GGA methodologies.  Therefore, {\it
  prima facie} it is not clear why there is a spread of predicted
polar fields since the polar field is a property of the \lao film
itself.  To shed some light on this issue of variations in the
theoretical predictions, we remark that it is possible to make an {\it
  a priori} estimate of the polar field in \lao using only bulk
properties of the material.  Denoting the lattice constant of \lao as
$a_{\rm LAO}$, the \lao atomic planes along (001) are alternately
(LaO)$^+$ and (AlO$_2$)$^-$ and separated by a distance of $a_{\rm
  LAO}/2$.  A simple parallel plate model can be made, as illustrated
in Fig. \ref{fig:average}. The $\pm e/a_{\rm STO}^2$ surface charge on
each plane is spread over a \sto unit cell area $(a_{\rm STO})^2$ and
the resulting electric fields are screened by the \lao dielectric
constant, $\epslao$.  Since the field only acts in half of each unit
cell, we have for the potential energy change $\Delta V$ across a unit
cell
\begin{equation}
\label{deltaV}
\Delta V = \frac{a_{\rm LAO}}{2}\cdot\frac{4\pi e^2}{\epslao (a_{\rm STO})^2}
\end{equation}
or converting to the average internal polar electric field
\begin{equation}
\label{averageE}
E_{polar} = \frac{2\pi e}{\epslao(a_{\rm STO})^2}
\end{equation}
Using experimental values of $a_{\rm STO}=3.905$ \AA, $a_{\rm LAO}=3.789$
\AA, we have 

\begin{equation}
\label{numericalVE}
\Delta V = \frac{22.5~{\rm eV}}{\epslao}\ \ \ , \ \ \ 
E_{polar} = \frac{5.94~{\rm V/\AA}}{\epslao}\,.
\end{equation}
Therefore, each calculation of $\Delta V$ or $E_{polar}$, and 
thus the resulting critical separation, can be directly 
mapped to a value of $\epslao$ implicit in the calculation.

\begin{figure*}[t!]
\includegraphics[angle=-90, width=15cm]{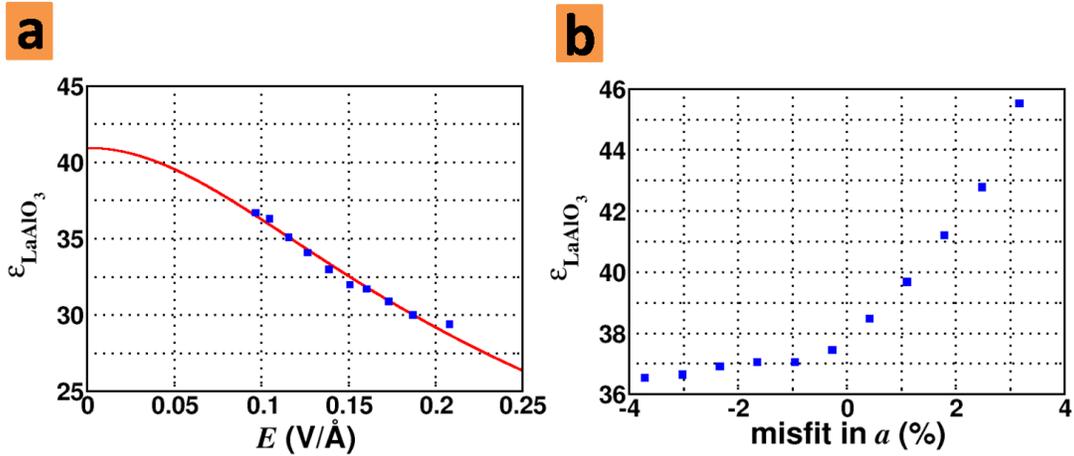}
\caption{\label{fig:diaelectric} \textbf{a)} Electric field dependence of 
the bulk \lao 
dielectric constant. The solid line is a fit to the first principles 
calculation, using the Landau-Devonshire formalism Eq.~(\ref{Landau}) 
\cite{Devo-Phil-1949} with $\epsilon_0=40.95$ and $E_0=0.15$ V/\AA. 
\textbf{b)} Strain 
dependence of the \lao dielectric constant. The internal electric 
field is set to be 0.026 V/\AA~(weak enough that the \lao remains insulating). 
 The $c/a$ ratio is optimized for each $a$.  
The strain is measured with respect to the theoretically predicted lattice 
constant of \sto (in experiments the \lao film is coherently strained 
to the \sto substrate).}
\end{figure*}

\begin{table*}
\caption{Polar properties of \lao films and $n$-type interfaces 
from first principles calculations.  Values of the potential change
 per added \lao unit cell $\Delta V$ and the polar field $E_{polar}$ 
are related by Eq.~(\ref{numericalVE}). In each reference, only one 
of the two values is reported.  The same equation then provides the
 dielectric constant $\epslao$.}
\begin{center}
\begin{tabular*}{1.0\textwidth}{@{\extracolsep{\fill}}ccccc}
\hline\hline
$\Delta V$ (eV) & $E_{polar}$ (V/\AA) & Estimated critical separation & $\epslao$ & Reference\\
\hline
0.4 & 0.1 & $>$6 & 56 & Pentcheva and Pickett \cite{Pickett-PRL-2009}\\
0.6 & 0.16 & $>$4 & 38 & Cen \etal \cite{Cen-NatMat-2008}\\
0.7 & 0.19  & 6 & 32 & Chen \etal \cite{Chen-PRB-2009}\\
0.8 & 0.2 & 5 & 30 & Son \etal \cite{Son-PRB-2009}\\
0.91 & 0.24 & 4 to 5 & 25 & Lee and Demkov \cite{Demkov-PRB-2008}\\
\hline\hline
\end{tabular*}
\end{center}
\label{tab:polarfield}
\end{table*}
In Table~\ref{tab:polarfield}, we compile the published values of 
$\Delta V$ or $E_{polar}$ and back out $\epslao$ as per 
Eq.~(\ref{numericalVE}).  
As expected, larger $\Delta V$ or $E_{polar}$ lead to smaller predicted critical
 separations, as well as smaller implied $\epslao$. The range of $\epslao$ 
spans a factor of more than two, which raises the question of why the 
spread is so large. 

The actual value of the dielectric constant of \lao 
depends on the strain state and internal electric field in the material, 
as well as technical choices such as the pseudopotential and basis set 
(a similar dependence study of $\epsilon_{\rm STO}$ is presented in 
\cite{Antons-PRB-2005}). We present the 
field and strain dependence of $\epsilon_{\rm LAO}$ for bulk \lao in 
Fig.~\ref{fig:diaelectric}a and \ref{fig:diaelectric}b. 
The field dependence is calculated in a slab geometry because 
only thin films can accommodate large internal electric fields 
of around $0.2$ V/\AA\ without becoming metallic. The dependence can 
be described phenomenologically using the Landau-Devonshire formalism 
\cite{Devo-Phil-1949} (the solid line in Fig. \ref{fig:diaelectric}a), 
which yields, approximately:

\begin{equation}
\label{Landau}
\epsilon(E) \simeq \epsilon_0\left(1+\left(\frac{E}{E_0}\right)^2\right)^{-1/3}
\end{equation}
where $\epsilon_0$ and $E_0$ are two fitting parameters. 

In addition,
$\epsilon_{\rm LAO}$ is affected by the strain state of the \lao
film, which is in epitaxial relation to the \sto substrate. The strain
dependence versus in-plane lattice constant is calculated using the Berry
phase method \cite{Ivo-PRL-2002} with a weak internal electric field
($0.026$ V/\AA)~so that the system remains insulating and the
polarization as a response to the external electric field is in the
linear region.  The $c$-axis is optimized for each choice of $a$-axis
parameter.  The data in Fig.~\ref{fig:diaelectric} show that 
$\epsilon_{\rm LAO}$ is sensitive to the in-plane strain and the
internal electric field of the \lao film. Both properties depend 
on technical choices: \eg, different choices of basis sets and/or 
pseudopotentials change the lattice constant of \sto by $\pm$ 1-2\%, 
which appears to be a minor difference but in fact has a significant 
effect on $\epsilon_{\rm LAO}$.  
We believe that a large part of the
theoretical spread is caused by these sensitivities.

To summarize, theoretical work to date has shown that the polar
catastrophe mechanism does indeed predict a critical separation for
$n$-type and $np$-type interfaces. Furthermore, reasonable values
of 5-6 unit cells of \lao are found. The well known
underestimation of band gaps from LDA or GGA means that the actual 
critical separations must be larger than the raw values produced 
by such approaches. That the predicted critical separations 
are larger than those found in most, but not all, experiments 
suggests that other mechanisms may play a role 
before or in conjunction with the polar catastrophe.

\begin{figure*}[t!]
\includegraphics[angle=0, width=11cm]{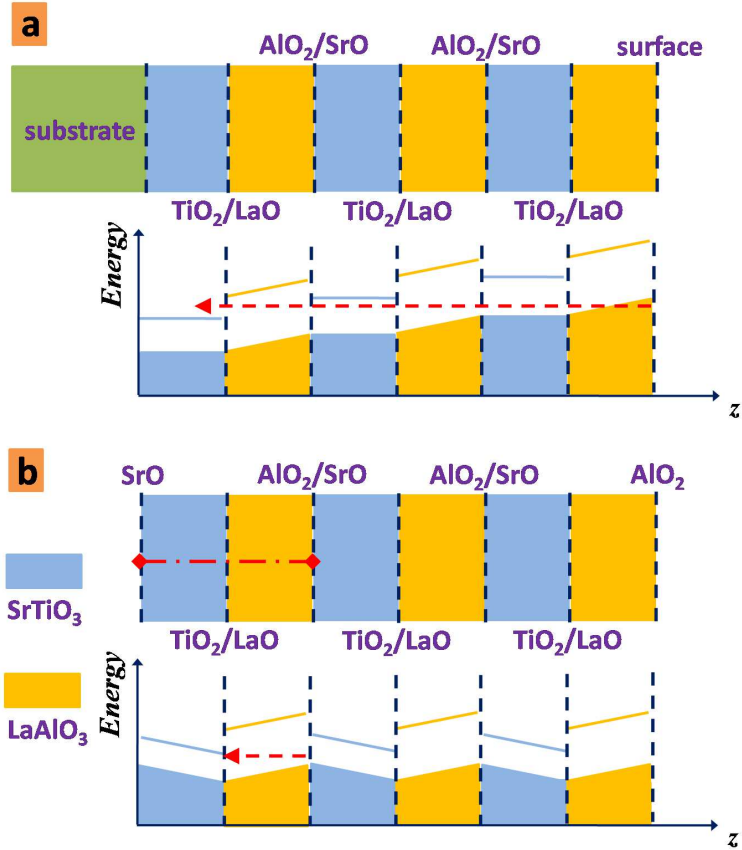}
\caption{\label{fig:np-combined} \textbf{a)} The experimental stoichiometric 
$np$-type superlattice and its corresponding energy diagram. The electrostatic 
potential is stair-like. 
\textbf{b)} The theoretical stoichiometric $np$-type superlattice with 
periodic boundary conditions and no vacuum region,  
and its corresponding energy diagram. Due to periodic 
boundary conditions, the potential has a
zigzag shape. The unit cell for the periodic boundary
conditions is illustrated by the red dot-dash line. In both cases, 
dashed lines with arrows show the direction of electron transfer when the 
polar catastrophe occurs in each system.}
\end{figure*}

Before concluding the discussion of the critial separation, we mention
that an insulating-to-metallic transition is also recently found in
the simulations of $np$-type superlattices with no vacuum regions
\cite{Bristowe-PRB-2009}. We would like to point out that there is a
subtle difference between experimental and simulated $np$-type
superlattices (illustrated in Fig.~\ref{fig:np-combined}). For an
experimental $np$-type superlattice, the films are of finite size,
terminate at surfaces, and are not subject to periodic boundary
conditions.  For such systems, a ``global'' insulating-to-metallic
transition can occur, which is missing in simulations with enforced
periodic boundary conditions.  As per Fig.~\ref{fig:np-combined}a,
before any electronic reconstructions takes place, the energy diagram
of an experimental $np$-type superlattice is stair-like: the potential
is constant in the nonpolar \sto and increases linearly in the polar
\lao layers. When the superlattice is thick enough, the electrons
filling the \lao valence bands on the surface tunnel into the empty
conduction bands of \sto substrate. This is the ``global''
insulating-to-metallic transition as the transferred electrons and
holes generate an additional internal electric field throughout the
\textit{entire} superlattice.  This additional field counteracts the polar
field of \lao film and distorts the electrostatic potential, as
illustrated in Fig. \ref{fig:np-combined}b.  Thus the limit of 
infinitely thick
experimental $np$-type superlattices can be simulated by imposing
periodic boundary conditions imposed on one superlattice unit cell.
Separately, if one increases the thickness of the \lao film in each
periodic unit, there will be a ``local'' insulating-to-metallic
transition in which electrons in the \lao valence states tunnel into
the \sto conduction bands in the \textit{same} unit superlattice.  We
note that the ``global'' transition (in experiment) or the periodic
boundary conditions (in computation) generates a counteracting field
that polarizes the \sto and reduces the polar field in the \lao film.  This
explains the smaller value of 0.057 V/\AA
\cite{Bristowe-PRB-2009} obtained in this approach. In earlier
calculations using relatively small unit cells \cite{Freeman-PRB-2006,
  Albina-PRB-2007, Gemming-Mater-2006}, the theoretical $np$-type
superlattices turn out to be insulating at \lao thicknesses that
make the $np$-type interface metallic. By employing much thicker films
in the superlattice unit cell (12 u.c. of \sto and \lao) an
insulating-to-metallic transition is found \cite{Bristowe-PRB-2009},
and as expected from the above reasoning the critical separation of 
$np$-type superlattices is quite a bit larger than 
that of $np$-type interfaces.
    
\subsubsection{Orbital character of interface bands}

In both simulations \cite{Pickett-PRB-2006,
Satpathy-PRL-2008, Chen-PRB-2009, Son-PRB-2009} and
experiments \cite{Salluzzo-PRL-2009}, it has been seen that the 
Ti-derived $t_{2g}$ degeneracy is split at the $n$-type interface. 
Theoretical studies show that the lowest occupied bands 
at the interface reside primarily in the \sto and have a clear 
Ti $d_{xy}$ character.  With increasing energy (or equivalently, 
sheet carrier density), $d_{xz}$ and $d_{yz}$ derived states also 
become partially occupied.  Therefore, for all but the very lowest 
sheet carrier densities, multiple interfacial bands, with possibly 
different orbital character, are occupied.

Detailed band structures for the $n$-type interface have been 
presented  in Refs.~\cite{Satpathy-PRL-2008, Son-PRB-2009} 
with some differences. Popovi\'{c} \etal \cite{Satpathy-PRL-2008} 
use a symmetric double $n$-type superlattice, which enforces the 
presence of precisely $0.5e/a_{\textrm{STO}}^2$ at each $n$-type interface 
(equivalent to what would be found for a very thick \lao film). 
Son \etal \cite{Son-PRB-2009} use a stoichiometric $n$-type 
interface calculation and present the band structure for 5 
u.c. of LaAlO$_3$, which gives rise to $0.2e/a_{\textrm{STO}}^2$ 
at the interface. 
The energy difference between 
the lowest and next lowest interface band edges, both of $d_{xy}$ character,
is 0.1 eV when $0.2e/a_{\textrm{STO}}^2$ is present at the interface, 
but this value triples to 0.3 eV when $0.5e/a_{\textrm{STO}}^2$ is transferred.  
In other words, there is an evolution of the energy spacings of 
interface states as a function of electron doping.  Both works 
show that the lowest $d_{xy}$ band is localized in
 the (001) direction and thus can be said to have a strong 
two-dimensional (2D) character.  Higher energy interface bands 
extend further into the \sto substrate in the (001) direction 
and are thus more 3D in character.

\begin{figure*}[t!]
\includegraphics[angle=-90, width=16cm]{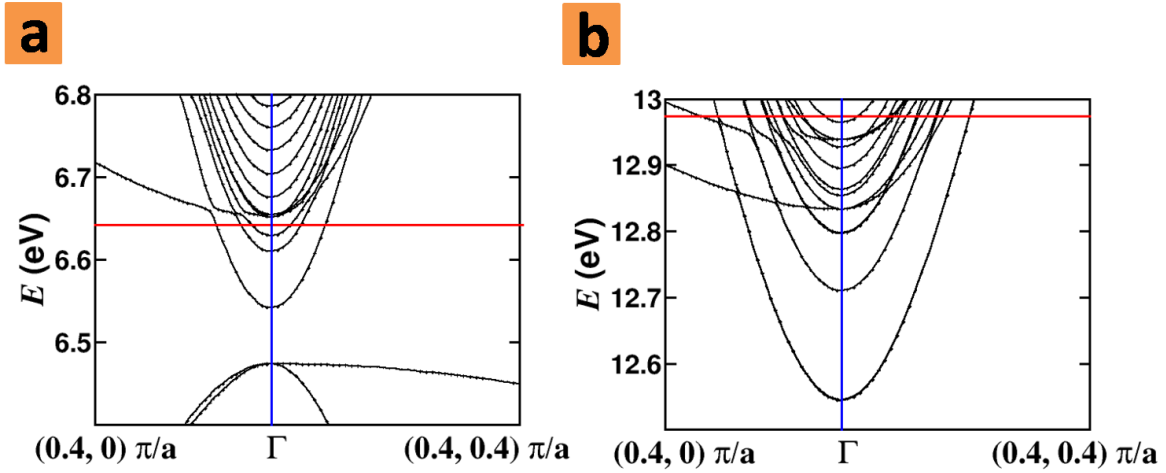}
\caption{\label{fig:bands} The band structures of an $n$-type interface 
and an $n$-type superlattice from DFT-LDA calculations. The solid horizontal red line is the Fermi level. 
\textbf{a)} $n$-type interface with 11 u.c. of 
\sto and 5 u.c. of \lao. The thickness of vacuum separating periodic copies 
is more than 30 \AA. The sheet carrier density is $0.1 e/a_{\textrm{STO}}^2$. 
\textbf{b)} $n$-type symmetric superlattice with 12.5 u.c. of 
\sto (with extra TiO$_2$ plane) and 4.5 u.c. of \lao (and extra LaO plane). As described in the text, the sheet carrier density is forced to be 
$0.5 e/a_{\textrm{STO}}^2$.}
\end{figure*}

A convenient way to summarize some of these facts is to show 
the Fermi surface of the interface system in the $k_x,k_y$ 
interface plane.  Fig. \ref{fig:bands} shows such a diagram.  
Following 
the above discussion, the lowest energy $d_{xy}$ band has 
the largest occupancy and thus the largest Fermi surface.  
With increasing energy, the bands have smaller Fermi surfaces.

\subsubsection{Possibility of Anderson localization}
\label{Anderson}

Discrepancies between theoretical predictions and experimental
measurements of the sheet carrier density \cite{Mannhart-Science-2006}
have stimulated discussion as to whether all of the transferred
electrons contribute to the
conductivity \cite{Satpathy-PRL-2008,Son-PRB-2009}.  An 
appealing argument for carriers to have different mobilities,
first proposed by Popovi\'{c} \etal \cite{Satpathy-PRL-2008}, is
based on the fact that, formally, all electronic states in 2D are
Anderson-localized by disorder in the thermodynamic limit.
Since contributions of localized states to transport are greatly 
suppressed, the idea is that electrons occupying 2D-like states 
should have a stronger tendency to localize and therefore not 
contribute to transport.

Popovi\'c \etal argue that the lowest $d_{xy}$ band, which is
strongly bound to the interface, should therefore be ignored in
predictions of the transport carrier density.  Furthermore, they
contend that the higher energy bands with $d_{xz}$ or $d_{yz}$
character should also localize because they have large effective 
masses along the (100) or (010) directions.  The only
electronic states that should contribute to transport are the 
higher-lying $d_{xy}$ bands.  They determine the carrier density
from these bands to be $8\times10^{13}$ cm$^{-2}$, which is closer to, 
but still a few times larger than, the
carrier density measured by Thiel \etal \cite{Mannhart-Science-2006}.
The results of Huijben \etal \cite{Huijben-NatMat-2006}, however, show a 
{\em larger} value of $1.4\times 10^{14}$ cm$^{-2}$.

Son \etal \cite{Son-PRB-2009} take this analysis one step further. They 
exclude the lowest $d_{xy}$, $d_{xz}$, and $d_{yz}$
bands as well as any $d_{xy}$ bands whose wave function is 
concentrated within $\sim 2$~nm of the interface, under the assumption 
that such electrons are easily scattered by interface roughness and 
therefore have low mobility.  Only electrons occupying loosely
 bound or unbound $d_{xy}$ bands are counted as contributing to 
transport. They find a sheet carrier density that
agrees well with the transport data from Ref.~\cite{Mannhart-Science-2006} 
(but not that from Ref.~\cite{Huijben-NatMat-2006}) 
in that it has a discontinuous jump at a critical thickness 
and then quickly saturates to a value of $\sim~2\times10^{13}$ 
cm$^{-2}$.  

We suggest that a direct experimental test of such 
localization would be to deliberately modify the amount of disorder by
growing the same interface system with variable quality.
Localization would show up as a variation in the
transport carrier density.  In the absence of such data,
however, we attempt to estimate theoretically the necessary 
conditions for the likelihood of the Anderson localization 
scenario at the non-interacting, one-particle level.  
We follow the review by Lee and Ramakrishnan
\cite{Lee-Ramakrishnan-RMP-1985}.

In 2D, all eigenstates of an infinitely large disordered system are
localized: any eigenfunction $\psi(\textbf{r})$ will
behave as $|\psi(\textbf{r})|\sim\exp(-|\textbf{r}-\textbf{r}_0|/\xi)$, 
where $\textbf{r}_0$ is the site about which the state is localized 
and $\xi$ is the localization length.  To understand the consequences 
of this for a finite system, the scaling theory of
localization \cite{Abrahams-scaling-1979} describes the
evolution of the sample conductance, $G(L)$, as the length scale 
$L$ evolves from microscopic to macroscopic dimensions. 
Localization suppresses the conductance, so that when
$L>\xi$,
\begin{equation}
G(L) \sim \frac{e^2}{\hbar}\exp(-L/\xi).
\end{equation}
For \stolao interfaces, the theoretical asymptotic sheet
density is $0.5e/a_{\textrm{STO}}^2$, or $3.3\times10^{14}$ cm$^{-2}$.  In order
to match transport experiments that give $2\times10^{13}$
cm$^{-2}$, $\sim$90\% of the electrons must be localized.  
Their contribution to the conductance must be at least
100 times smaller than the remaining $\sim$10\%.  Thus, a 
conservative bound is $\exp(L/\xi)>100$ or
$L/\xi>4.6$.  In a typical experiment, the contacts used for 
transport measurements can be about 1 mm apart \cite{Mannhart-Science-2006}, 
setting another conservative upper bound of $L=1$~mm, meaning 
$\xi<0.2$~mm.

The scaling theory also provides estimates for $\xi$.  
For the weak scattering scenario, $\xi$ is of order
\begin{equation}
\label{eq:xiloc}
\xi \sim l \exp\left( \frac{\pi}{2} k_Fl\right)
\end{equation}
where $k_F$ is the Fermi wave vector of the electrons being
considered and $l$ is the mean free path.  (Weak scattering is 
defined as $k_Fl\gg1$.)  The DFT band structures in 
Fig.~\ref{fig:bands}, which are similar to the results 
of Refs.~\cite{Satpathy-PRL-2008,Son-PRB-2009}, provide us with $k_F$ for
each band. From Fig.~\ref{fig:bands}b, we can extract out that the 
lowest $d_{xy}$ band has $ k_F = 0.28\pi/a_{\rm STO}$, 
while the next higher energy $d_{xy}$ band has $k_F= 0.21\pi/a_{\rm STO}$ 
or 2.3 nm$^{-1}$ and 1.7 nm$^{-1}$,
respectively. Higher energy and less occupied bands have smaller
$k_F$.  These values, along with Eq.~(\ref{eq:xiloc}), allow us 
to convert the bound on $\xi$ to a bound on $l$.

If $l$ is approximately the same for all the interface bands, then
$\xi$ will be larger for the lower energy bands, which have the larger
$k_F$ and will progressively decrease for higher energy bands.  This
is the inverse of the trend needed for localization to hold.  Thus,
for localization to occur, $l$ must be significantly larger for the
higher energy bands.  Namely, since the bands become more delocalized
into the \sto substrate with increasing energy, the interior of the
\sto substrate would have to be of high quality while the part with a
few nm of the interface would have to be sufficiently disordered.

Following this logic and using the above numerical 
values in Eq.~(\ref{eq:xiloc}), we find the equivalent 
conditions that $l<3$ nm for the lowest band and $l<4$ nm for the higher
$d_{xy}$ bands. These values are small, considering the following standard 
estimation of mean free path:

\begin{equation}
\label{eq:meanfreepath} l\sim \frac{\mu_H \hbar k_F}{e}
\end{equation}
Taking a typical value of Hall mobility (see Table \ref{tab:transport})
$\mu_H\sim 10^3$~cm$^2$V$^{-1}$s$^{-1}$ and 
$k_{F}\simeq 2$~nm$^{-1}$, then we obtain $l \sim 100$~nm. Regardless, 
assuming that a reasonable 2D 
cross section $\sigma$ for a relevant defect is a few nanometers, 
the typical spacing of such defects in the interface
plane is $\sim\sqrt{l\sigma}$ and must also be on the order
of a few nm or less.

We briefly consider what defects could produce such
small $l$.  One may dismiss surface
steps or edge dislocations in the \sto substrate. The former are
typically hundreds of nm apart, and experiments indicate that they do
not hinder transport, while the latter have a
small areal density of $\sim10^{-8}$ nm$^{-2}$ \cite{Thiel-PRL-2009}. 
The presence of 
extended defects every few nm can be excluded, as they would be
visible in transmission electron 
microscopy (TEM) images of the interfaces.  This
means that point-like defects in
the immediate vicinity of the interface are the most realistic 
possibility, and candidates include cation intermixtures and oxygen vacancies.


There is experimental evidence of La-Sr intermixture across
the $n$-type interface \cite{Hwang-NatMat-2006,Willmott-PRL-2007} 
on a scale of 1-2~nm.  Son \etal \cite{Son-PRB-2009} 
use this ``interface roughness'' as the basis for excluding
states localized within $\sim$2~nm of the interface.  However,
this roughness is in the direction normal to the interface, 
whereas the relevant spacing is that of defects in the interface 
plane.  For intermixtures to be responsible, La-Sr mixing must be 
rough over the scale of a few nm in the interface plane itself.  
Whether that occurs remains to be determined in future 
experiments and theoretical calculations.

Regarding oxygen vacancies, Nakagawa \etal \cite{Hwang-NatMat-2006}
infer a small density of oxygen vacancies (0.10$\pm$0.04
vacancies per 2D unit cell) at $n$-type interfaces. 
This defect density is of the right 
order of magnitude and might be considered a
viable candidate for the disorder mechanism. Whether these
oxygen vacancies have the necessary cross sections for scattering is a
separate question.  More importantly, oxygen vacancies at this
density can in principle donate $0.20e/a_{\textrm{STO}}^2$ ($1.3\times10^{14}$
cm$^{-2}$) to the interface, which is larger than the observed
transport density.  Even if the donation is imperfect, the number of
electrons being added to the already electron-populated interface would be
significant and these electrons would occupy high energy, loosely 
bound states. They would greatly {\it increase} the number of
electrons contributing to transport.

In summary, the Anderson localization scenario provides an attractive,
yet still unsubstantiated, picture to explain the low sheet carrier
densities observed in the transport experiments of Thiel \etal
\cite{Thiel-PRL-2009}.  A more quantitative analysis of the situation
leads us to the following conclusions: i) to induce the required
localization, the mean free paths of states strongly bound to the
interface must be smaller than those that are loosely bound; ii)
the mean free path in the interface plane for the strongly bound 
states could be at most a few nm; and iii) point-like defects are 
likely scattering centers, with cation intermixtures or oxygen
vacancies being possible candidates.  While theoretically feasible, 
the localization scenario requires that a number of factors work 
hand in hand.

We end by noting that this localization scenario \cite{Satpathy-PRL-2008} is
proposed to explain why theoretical values of sheet densities are
higher than those found in the experiments of Thiel \etal
\cite{Mannhart-Science-2006}.  On other hand, the sheet densities
measured by Huijben \etal \cite{Huijben-NatMat-2006} are larger in
magnitude, saturate to a value only a factor of two smaller than the
theoretical value of $0.5e/a_{\textrm{STO}}^2$, and 
show a monotonic increase with
\lao thickness, all in much closer agreement with existing
theoretical findings.

\subsubsection{Spatial extent of the electron gas}

As discussed in the review of experiments above, the experimental
consensus for the thickness of the electron gas at the $n$-type interface 
is that it is confined to a length scale on the 
order of nanometers in the samples grown at high 
oxygen partial pressure ($\sim 10^{-4}$ mbar) 
\cite{Mannhart-Science-2007, Basletic-NatMat-2008, Sing-PRL-2009}. For 
samples grown 
at low oxygen partial pressure ($\sim10^{-6}$ mbar), 
the conduction electron gas extends hundreds of $\mu$m into 
the \sto substrate 
\cite{Herranz-PRL-2007, Basletic-NatMat-2008}.  
In the latter case, the electrons may be 
due to extrinsic sources (\ie, oxygen vacancies) and thus not 
determined by the intrinsic properties of the interface.  
The main contribution of {\it ab initio} theory in this matter 
is to provide detailed information about the shape and 
distribution of the electron gas in the intrinsic limit.

Janicka \etal \cite{Tsymbal-PRL-2009} use double $n$-type 
symmetric supercells and find that the charge density
profile of the electron gas follows an exponential decay,
$e^{-z/\delta}$ (where the first TiO$_2$ layer is at $z=0$ 
and the \sto substrate is on the $z>0$ side). Fitting to DFT
data, they obtain $\delta\simeq1$~nm.   
Due to the imposed symmetry, the geometry used by Janicka 
\etal \cite{Tsymbal-PRL-2009} contains five 
\sto unit cells of substrate ($\sim$ 2 nm).  In our work 
\cite{Chen-PRB-2009}, we use a stoichiometric $n$-type interface 
in a supercell with 11 u.c. of SrTiO$_3$, and we find that the 
conduction electron density profile decays rapidly for the 
first few unit cells away from the interface, but has a longer, 
non-exponential tail further into the substrate.  
Son \etal \cite{Son-PRB-2009} have performed the largest 
simulations to date with 15 to 30 u.c. of \sto and also find
exponential decay close to the interface that turns into 
an algebraic decay further into the substrate (see Fig. 5 of that work).  
Specifically, their calculations show that within 4 nm of the interface, 
the charge density decays exponentially with $\delta = 1.84$ nm. 
Further into the \sto substrate, the charge density profile is 
found to decay more slowly with the approximate form $b/(z-z_0)$.

\subsubsection{Binding mechanism to the interface}

The confinement of the electron gases at interfaces is usually
attributed to band bending, a well-known concept in
metal-oxide-semiconductor (MOS) interfaces.  Janicka 
\etal \cite{Tsymbal-PRL-2009} find that the confinement of the quasi 2D
electron gas at the $n$-type interface of \stolao is not
solely due to band bending. They suggest that the formation of
metal-induced-gap-states (MIGS) due to the presence of the conduction
electrons at the interface plays an important role, allowing the
electron gas to extend further into the \sto substrate than
simple band bending would allow.  

An important question is why the electrons are bound 
to the interface in the first place. One answer 
is that, after the polar catastrophe takes place, the electrons at the 
$n$-type interface feel the electrostatic attraction from the holes on 
the other side of the \lao film, and this attraction binds them to 
the interface.  While this must play a role, it is not 
sufficient for the following two reasons.  First, for the 
double $n$-type symmetric supercells, which have no polar fields and 
where there are no holes present to create the electrostatic attraction, 
DFT simulations still find a bound electron gas at the interface.  
This has already been reported in 
Refs.~\cite{Tsymbal-PRL-2009,Satpathy-PRL-2008}. In fact, 
for symmetric supercells with no polar fields, one would expect the 
electrons to occupy the lower-energy conduction band edge of 
the two materials (\sto for this pair of materials) and to 
spread throughout the substrate.  Second, holes at the $p$-type interface 
should feel the same attractive force, but instead are found to delocalize 
into the \sto substrate.  We report this effect for stoichiometric $p$-type 
interfaces \cite{Chen-PRB-2009} and have verified that this 
holds for double $p$-type symmetric superlattices. Thus, the attraction 
of the opposite carriers across the \lao film is not sufficient to 
strongly bind them to the interface.  Furthermore, there is an asymmetry 
between $n$-type and $p$-type interfaces in terms of binding carriers 
that is intrinsic to the interface structure itself.

We have proposed an explanation for this asymmetry \cite{Chen-PRB-2009}.  
In both \sto and LaAlO$_3$, 
the conduction band edge stems from the transition metal $d$ orbitals.  
However, in \sto the Ti atoms are on the $B$ site while in \lao the La atoms
are on the $A$ 
site (in the standard $ABO_3$ notation).  Thus in the bulk phase of SrTiO$_3$, 
the nearest Ti atoms are one lattice constant apart, but at the 
$n$-type TiO$_2$/LaO interface the Ti-La distance is $\sqrt{3}/2$ times 
one lattice constant. 
This proximity, combined with the large spatial extent of the La $d$ orbitals, 
enhances the tight-binding hopping element between the two transition 
metals to the extent that it becomes about 100 times larger than the hopping 
element between Ti $d$ orbitals in bulk SrTiO$_3$. This large hopping matrix 
element creates a pair of bonding and antibonding states composed of 
superpositions of the Ti and La $d$ states right at the interface: the 
low-energy bonding state is primarily of Ti $d$ character and the 
anti-bonding one of La $d$ character. Compared to the other 
Ti atoms in the \sto substrate, the Ti at the interface are special, as they 
have a lower energy and are thus the most favorable sites for electrons to 
accumulate when the metal-insulator transition takes place.  
(As detailed further below, the $p$-type interface has no enhanced 
hopping, and thus has no such special binding mechanism.)

\begin{figure*}[t!]
\includegraphics[angle=-90, width=13cm]{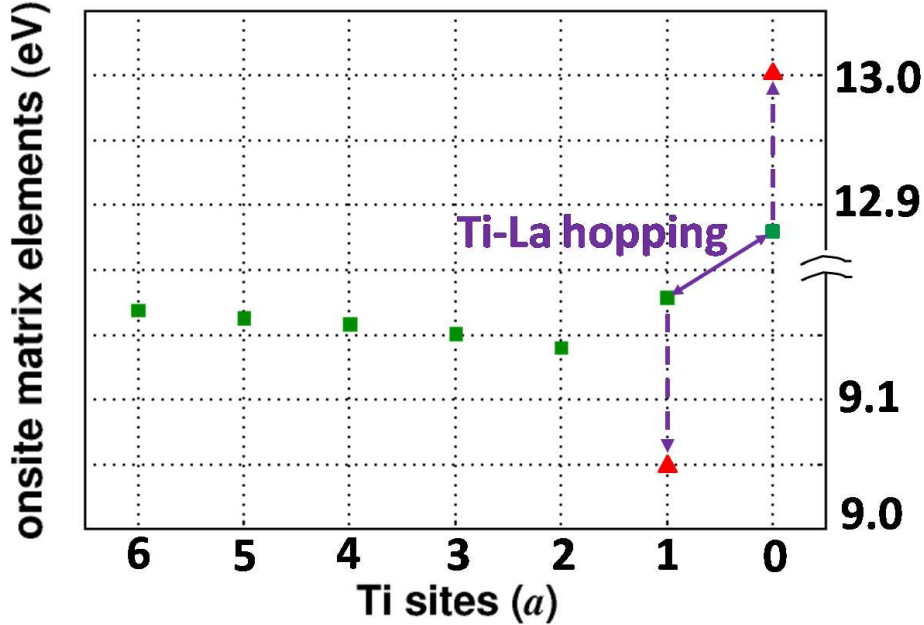}
\caption{\label{fig:onsite} Green squares show the on-site tight-binding 
matrix elements at the $n$-type interface computed from first principles 
for a simulation with 11 u.c. of \sto and 4 u.c. of \lao 
(the insulating-to-metallic transition occurs at this thickness in the 
DFT simulation). The on-site energies are for the Ti $d_{xy}$ and 
La $d_{xy}$ orbitals. The La atom is at site 0, and the Ti atoms are at 
sites 1-6. The $n$-type interface 
is located between sites 0 and 1.  Note that the first Ti atom (at site 1) does 
not have the lowest on-site energy, but has the highest electron 
density.  The red triangles show the effect of including the large hopping 
element between Ti and La at the interface. The La level is pushed up and 
the Ti level is pushed down, so inclusion of the hopping element now makes 
the first Ti the lowest energy site for electron accumulation.}
\end{figure*}

In Ref.~\cite{Chen-PRB-2009}, we verify this explanation by mechanically 
moving the Ti at the interface to elongate (weaken) or shorten (strengthen) 
the Ti-La bond.  As expected, when the Ti-La bond is strengthened, 
electrons more strongly localize towards the interface, and when the bond 
is weakened, electrons move farther into the substrate.  
As an alternate check, we compute the on-site energies of the 
transition metal $d_{xy}$ orbitals from first principles in the 
interface region for a 4 u.c. thick \lao film that has undergone 
the polar catastrophe (within LDA). The results are shown in Fig. 
\ref{fig:onsite}. The on-site energy for atom 
$i$ is the diagonal matrix element 
$\langle d^i_{xy}|\hat H|d^i_{xy}\rangle$, where $\hat H$ is the 
Kohn-Sham Hamiltonian and $d^i_{xy}$ is the atomic orbital at 
that site.  As expected, the La $d_{xy}$ is at higher energy 
than the Ti $d_{xy}$ due to the conduction band offset.  
The salient point of the figure is that the on-site energy 
of the Ti atom right at the interface is not the lowest one,  
but simulations show that the electron density is highest 
on the first interfacial Ti atom at the interface.  
The resolution to the puzzle is to 
include the large Ti-La hopping element of 0.7 eV. As indicated 
in Fig. \ref{fig:onsite}, this pushes up the La level and lowers the Ti 
level at the interface.  With this hopping included, the Ti at 
the interface is now the most favorable site for binding electrons and 
should accumulate the largest number of electrons.

These findings imply that the spatial extent of the electron gas 
is determined in large part by the Ti-La coupling at the interface.  
In particular, one does not expect to find an equally 
strongly bound electron gas 
when \lao is replaced by another perovskite in which the transition metal 
is either on the $B$ site or has more localized atomic
orbitals.  An experimental test of this prediction is 
straightforward in principle and requires epitaxial 
growth of other oxides on \sto and the same types of measurements 
already performed for \stolao interfaces. If this physical 
picture is correct, one can consider a number of possibilities for
 engineering the spatial profile and properties of the electron gas.  
Chemical substitutions right at the interface will change both the 
spatial extent of atomic orbitals as well as the interatomic spacings 
and thus modify the tight-binding hopping matrix elements.  
Alternately, mechanical motion of the transition metal atoms 
might be achieved by using piezeoelectric or ferroelectric 
substrates that will in turn change the hopping elements 
through direct modifications of inter-atomic distances.  

\subsubsection{Magnetic ordering}

As discussed above, experiments have shown intriguing 
magnetic properties at the \stolao interface. Several groups have 
attempted to tackle the question of magnetic ordering at the interface 
theoretically with the LDA+U approach 
\cite{Pickett-PRB-2006,Pickett-PRB-2008, Kelly-EPL-2008, Tsymbal-JAP-2008}. 

Pentcheva and Pickett studied both unrelaxed \cite{Pickett-PRB-2006} 
and relaxed \cite{Pickett-PRB-2008} $n$-type interfaces using a
$c(2\times2)$ interface unit cell with $U=8$~eV and $J=1$~eV. 
They report that atomic relaxation does not play an important role in
determining the magnetic states and their energetic orderings. 
The ground state in both cases is a ferromagnetic insulating state 
with disproportionated charge ordering in a checkerboard pattern 
in the plane of the interface. Perpendicular to the interface, they find 
that the excess 0.5 electrons per 2D cell are localized and 
reside in a single atomic plane right at the interface.  
Without the Hubbard $U$ \cite{Pickett-PRB-2008}, 
they find that the
ferromagnetic state and charge ordering are suppressed, that the 
interface is metallic, and that the excess electrons are 
delocalized over several \sto unit cells.

Kelly \etal \cite{Kelly-EPL-2008} use the
rotationally invariant LDA+U method  
\cite{Lie-PRB-1995} with $U=5$ and 
$J=0.64$ eV.  In contrast to the results
of Pentcheva and Pickett, they find that the
atomic geometry, in particular a GdFeO$_3$-type distortion, is
crucial in reducing the bandwidth of occupied Ti $d$ states and
in stabilizing magnetic orderings.  Their ground state is a 
$p(2\times2)$ anti-ferromagnetic insulating state with charge ordering.  
However, they point out that by imposing
ferromagnetic ordering, they obtain a $c(2\times2)$ insulating state
with charge ordering, similar to the finding of Pentcheva and Pickett, 
which is 10~meV higher than their anti-ferromagnetic ground state. 
However, this energy difference is small, close to the limit of 
their calculation accuracy \cite{Kelly-EPL-2008}.  

Janicka \etal \cite{Tsymbal-JAP-2008} use a $(1\times1)$
interface unit cell and obtain different results. They find that 
when the excess electrons are forced to be confined to within 
1.5 unit cells of \sto at the interface, the ground state is a 
metallic ferromagnet, even in the absence of a Hubbard $U$ correction. 
By increasing the \sto
thickness, they observe a decreasing magnetic moment and
disappearance of the ferromagnetic state for 4.5 unit cells of
SrTiO$_3$. Upon applying $U$=5 eV, they find that 
ferromagnetism is stablized in the ground 
state for all the thicknesses (up to 4.5 \sto unit cells). 
However, their interfaces remain
metallic, in contrast to the above results. One explanation is 
that a $(1\times1)$ unit cell is too small to allow 
for charge ordering --- all Ti sites are forced to be equivalent, 
the eigenstates must be partially filled extended Bloch 
states, and one thus expects to find  metallic behavior.

Charge ordering is an important ingredient in determining 
the correct magnetic ground state of the $n$-type interface, and the 
studies above demonstrate that accurate simulations require supercells 
with sufficient lateral extent to allow for  
this phenomenon.  We end this section by noting that, while the LDA+U 
approaches to date have found magnetic states at the interface, 
the calculations 
with larger unit cells find insulating ground states, in contrast to 
experiments that report mobile electrons even 
at the lowest of temperatures. 

\subsection{The $p$-type interface}

A number of first principles calculations have attempted to shed 
light on the insulating nature of the $p$-type interface as 
well as its general physical properties. For the insulating 
behavior, both intrinsic and extrinsic mechanisms have been investigated.  
Concomitant with the smaller number of experimental studies of 
this interface, there has also been less theoretical work 
on this system.

\subsubsection{Intrinsic $p$-type interface: DFT results}

Theoretical calculations that have used DFT within the standard 
LDA or GGA approach report a metallic state at the 
$p$-type interface \cite{Pickett-PRB-2006,Freeman-PRB-2006,Chen-PRB-2009}.  
The holes at the interface reside on the 
perovskite oxygen sublattice, their electronic states have 
essentially pure oxygen $p$ character as expected for the 
valence band edge of a perovskite, and they occupy the top 
portion of the valence band, creating a Fermi surface.  
At this level of description, the metallic character is 
clearly at odds with experiments.

As mentioned above, the holes at the $p$-type interface 
are not strongly bound to the interface but instead delocalize 
into the \sto substrate.  We have found this to hold for both 
stoichiometric $p$-type interfaces \cite{Chen-PRB-2009} as well 
as large double $p$-type symmetric supercells. 
Unlike the $n$-type interface, where binding is enhanced by the 
unusual Ti-La tight-binding hopping element, there appears to be 
no such binding force present for the $p$-type interface. The oxygen 
sublattice is continuous and unmodified across the $p$-type interface.  
Therefore, one does not expect any special hopping matrix element to be 
present, and calculations report that the oxygen-oxygen 
hopping elements are essentially constant and unchanged across 
the interface. The polar field in the \lao will drive holes 
out of the \lao film and into the \sto film at the $p$-type interface, 
but once in the SrTiO$_3$, the holes do not have 
any reason to be bound to the interface.  

In our previous work \cite{Chen-PRB-2009}, we investigate supercells with 
up to 11 u.c. of SrTiO$_3$, and find that the holes are delocalized 
within the \sto substrate for such thicknesses. The holes 
do feel the attractive electrostatic force from the electrons on the 
other side of the \lao film.  Therefore, one expects that the holes 
should be bound to the interface for thicker \sto films.  What the 
calculations show is that the length scale for the hole binding is 
larger than 11 \sto unit cells, meaning that 
the binding energies will be weaker as well (when compared to the 
electrons at the $n$-type interface). At the DFT level, 
one expects an unbound or very weakly bound hole gas for the $p$-type 
interface. However, the DFT prediction of a metallic hole gas will 
still hold for such an ideal interface. For a more complete picture 
of how insulating behavior might arise in this picture, see the 
section \ref{ov at ptype} on oxygen vacancies further below.

\subsubsection{Intrinsic effects at the $p$-type interface: the hole polaron}

Pentcheva and Pickett \cite{Pickett-PRB-2006} investigate the 
possibility that strong correlation effects can induce a self-trapped 
hole polaron at the $p$-type interface.  To this end, they use an 
LDA+U approach with $U$=7~eV on the oxygen $p$-states. They 
study a symmetric double $p$-type superlattice with the atoms 
fixed in the ideal perovskite positions and a $p(2\times2)$ 
interface unit cell. They consider both ferromagnetic and anti-ferromagnetic 
orderings. The ferromagnetic state they find exhibits disproportionated 
charge ordering and is half-metallic, at variance with the insulating 
behavior observed in experiments. They find the antiferromagnetic 
state to be charge ordered and insulating, with a very small 
energy gap of 50~meV. The half-metallic ferromagnetic state is 
energetically favored by 0.15~eV over the insulating anti-ferromagnetic state. 

These results provide some evidence that correlation effects missing in 
LDA or GGA approaches can create interface electronic structures that 
differ from the simple metallic prediction of the LDA/GGA. 
With increasing $U$, charge-ordered localized states 
become energetically favored due to the increasing cost of double 
occupancy on an atomic site.  Therefore, one expects a transition 
from the delocalized, metallic bands of the LDA/GGA prediction to 
localized states upon increasing $U$. Open questions are how 
sensitive these results are to the particular $U$ value chosen 
and what the ``correct'' value of $U$ may be.  Separately, the 
above findings are based on unrelaxed atomic structures, 
and it is unclear whether atomic relaxation would favor antiferromagnetism.  

\subsubsection{Oxygen vacancies at the $p$-type interface}
\label{ov at ptype}

Extrinsic effects such as those from defects or impurities may also 
help explain the insulating behavior of the $p$-type interface.  
Oxygen vacancies are one of the most prevalent 
defects in the \sto substrate, and thus a number of research groups 
have focused on studying their interactions with the $p$-type 
interface.  That oxygen vacancies can cause the insulating behavior 
was first suggested by Nakagawa \etal \cite{Hwang-NatMat-2006}. 

Pentcheva and Pickett \cite{Pickett-PRB-2006} consider unrelaxed 
$p$-type superlattices with 25\% of oxygen vacancies (\ie, one
oxygen vacancy per four 2D unit cells) in either the SrO or
AlO$_2$ atomic planes at the SrO/AlO$_2$ $p$-type interface and find 
that the ground state is insulating and nonmagnetic in both cases. 
However, the Fermi level is found to be located in a narrow dip 
of the density of states, and the energy gap is too small to distinguish. 
(When examining oxygen vacancies, Pentcheva and Pickett do not apply 
the Hubbard $U$ that they use when investigating the hole polaron.)

Park \etal \cite{Freeman-PRB-2006} studied both unrelaxed and
relaxed double $p$-type symmetric superlattices 
without Hubbard $U$. Upon allowing relaxation and including 
oxygen vacancies in
the SrO plane at the interface, which favors oxygen
vacancy formation over the AlO$_2$ layer,
they find that the $p$-type superlattice with 25\% of oxygen 
vacancies remains metallic, in contrast to the prediction of the 
simple ionic limit. However, increasing the oxygen vacancy concentration 
to 50\% results in insulating behavior. They suggest that strong 
hybridization between Ti $d$-orbitals and O $p$-orbitals close to 
the Fermi level makes it difficult for the two electrons of an oxygen 
vacancy to easily compensate the holes.  This result is in 
qualitative agreement with the experimental finding that 32\% oxygen 
vacancies per 2D unit cell (and not 25\%) are found at the insulating
$p$-type interface \cite{Hwang-NatMat-2006}.

The above two studies employ symmetric superlattices and consider 
oxygen vacancies directly at the $p$-type interface.  Our work on 
stoichiometric $p$-type supercells with a $p(2\times2)$ interface unit 
cell \cite{Chen-PRB-2009} investigates the energetics of oxygen vacancy 
formation in the atomic planes at and away from the $p$-type 
interface.  We find that, for a 25\% oxygen vacancy concentration, 
the formation energy decreases as the oxygen vacancies move away 
from the $p$-type interface.  The convergence to the bulk \sto 
oxygen vacancy formation energy is achieved at a distance of 
three unit cells from the interface.  These results show that 
the $p$-type interface repels oxygen vacancies, which 
instead prefer to stay in the bulk of the \sto substrate.  

Since both holes and vacancies are repelled by the interface, 
these results suggest that the interface region is not only 
defect free but also free of holes. Therefore, the interface itself will 
be insulating. The remaining issue is to understand why the 
holes do not contribute to transport.  Based on the picture above, 
we propose the following qualitative possibility:  the \sto 
substrate will inevitably have some oxygen vacancies, either due 
to growth conditions or thermal fluctuations.  As long as the \sto 
substrate is thick enough to contain at least one oxygen vacancy 
per four 2D unit cell (a very low 3D concentration for a thick substrate), 
the holes can become trapped (\textit{i.e.} annihilated) by the two electrons 
of the oxygen vacancy and thus will not contribute to transport.  
This hypothesis could be tested by large scale simulations that 
contain $p$-type interfaces, oxygen vacancies, and relatively thick \sto slabs.

\section{Outstanding puzzles}
\label{puzzles}

Despite experimental and theoretical efforts, the physical
origins of some aspects of the \stolao interface remain
undetermined. We have mentioned several of these questions in the course
of this progress report. In this section, we discuss them in more
detail and suggest several experiments and computations that may help
to further illuminate the underlying physics of the range of
intriguing behavior observed in this remarkable system.

\subsection{Whence insulating $n$-type interfaces?}
\label{insulating n-type}

It is found in experiment that for samples grown at very high oxygen 
partial pressure ($> 10^{-2}$ mbar) \cite{Kala-PRB-2007} or 
annealed at 300 mbar during cooling \cite{Herranz-PRL-2007}, 
the $n$-type interfaces can become completely insulating. 
In fact, the conducting properties of $n$-type interfaces become poorer 
with increasing \poo\ during the 
growth process (from 10$^{-6}$ to 10$^{-3}$~mbar) \cite{Herranz-PRL-2007}.  
With presumably better 
stoichiometry control, one would expect that the transport properties
 would converge to those of an ideal interface, where the 
polar catastrophe plays the dominant role. 
What is noticeable is that the high-pressure growth experimental results are at 
variance with these expectations.

At first glance, this phenomenon would imply that the observed 
interface conductivity comes only from the oxygen 
vacancies. With the higher \poo\ or post-annealing to repair 
the oxygen off-stoichiometry, the concentration of oxygen 
vacancies should decrease, resulting in higher sheet 
resistance and lower sheet carrier density. However, the 
sheet carrier density does show an abrupt jump at 4 unit 
cells of \lao \cite{Mannhart-Science-2006}, which 
is difficult to explain using oxygen vacancy doping alone. 

We mention here a number of possible scenarios and related 
questions. The first is that, for ideal $n$-type interfaces 
and for polar LaAlO$_3$, the polar catastrophe does not
take place and that the theoretical models are 
unable to describe the correct 
physical properties. While logically possible, this scenario seems unlikely
because the polar catastrophe for idealized interfaces is based on 
simple physical properties such as ionic charges of the cations. 
Next, one can 
consider the possibility 
that the theoretical results are correct but inapplicable;  
that the \lao becomes non-polar when the samples are grown 
under high \poo\ or post-annealed.  Perhaps, aside from 
removing oxygen vacancies, the oxidation has other side 
effects.  The oxidation introduces oxygen atoms at the 
surfaces of the samples which then diffuse into the body 
of the materials. Does a larger flux of oxygen through the 
lattice affect the cation sublattice by enhancing disorder 
and/or intermixing? Does one achieve sharp and 
ideal interfaces or is the disorder sufficient to make the \lao nonpolar?

Recent work by Segal \etal on molecular beam 
epitaxy grown \stolao interfaces yield a number of interesting findings 
concerning some of these questions.  Similar to 
Thiel \etal \cite{Mannhart-Science-2006}, a critical separation of 
4 unit cells of \lao is found, as well as similar sheet carrier 
densities in transport.  However, x-ray spectra of the La core states 
do not reveal the expected \lao polar field.  Quantitatively, the 
polar field is $\sim$10 times smaller than the typical theoretical 
value of $\sim$0.2 eV/\AA\ for samples with \lao thicknesses below and above 
the critical separation.  This provides some evidence that the \lao
 may not be polar in some samples.  However, it also creates 
puzzles:  if the \lao is not strongly polar, then what mechanism 
leads to a critical thickness close to the 
expected value?

One direct experimental probe for these questions 
is the electron energy loss spectroscopy (EELS) performed in the 
scanning transmission electron microscope (STEM). 
Detailed studies of the Ti, La, and O core 
level have provided valuable information on the distributions of the 
cations and oxygen vacancies at $n$-type and $p$-type interfaces 
\cite{Hwang-NatMat-2006}.  However, we are not aware 
of data published to date on the La core levels for relatively thin \lao 
films close to the critical thickness.  The existence of the polar 
field in \lao will lead to an energy shift between neighboring 
La core levels along the (001) \lao film direction.  An expected 
shift of about 0.7 eV per \lao unit cell is within present experimental 
resolution.  That will allow for direct visualization of the strength 
of the \lao polar field.

On the theoretical front, because ideal interfaces and films may not
 provide a complete explanation, first principles studies 
can investigate aspects of disorder and intermixing.  Specifically, 
which types of cation intermixtures are thermodynamically favored?  
What effect do they have on the polar field (if any)?  Do they modify
 the critical separation or create new mechanisms for electron doping 
independent of the polar catastrophe?  In addition, any experimental 
sample will have some degrees of off-stoichiometry.  Theoretical 
studies of the possible importance of this disorder effect would 
be valuable.

\subsection{What is the value of the critical separation?}

As discussed above, first principles simulations predict 
an $n$-type interface critical separation of $5-6$ unit cells of LaAlO$_3$. 
Recently, based on evidence from 
optical second harmonic generation,  Savoia \etal \cite{Savoia-PRB-2009} 
propose that, even at 3 unit cells of 
LaAlO$_3$, electrons are injected into the interface but become 
localized. The conductance does not appear until the electron gas 
becomes more uniform at a larger \lao thickness. Sing \etal 
\cite{Sing-PRL-2009} use hard x-ray photoelectron 
spectroscopy and find finite sheet carrier density at samples 
with 2 u.c. of LaAlO$_3$. They remark that part of the carrier 
concentration may be of intrinsic origin, but it does not lead 
to conduction. 

What are some possible mechanisms that lead to a smaller 
critical separation than the polar catastrophe would predict? 
First, oxygen vacancies in the \sto substrate are an 
extrinsic doping mechanism that is unlikely here because it  
can not account simply for the abrupt insulating-to-metallic 
transition. Cation substitution \cite{Willmott-PRL-2007} 
which results in the formation of metallic La$_{1-x}$Sr$_{x}$TiO$_3$ 
is another 
possible mechanism. However, some
details are not clear yet for this mechanism. What is the most stable 
distribution of cation substitution? How does cation substitution 
depend on the thickness of LaAlO$_3$? In the very thin \lao films 
($< 4$ unit cells), is the cation mixing suppressed or enhanced? 
Cen \etal \cite{Cen-NatMat-2008} have argued that the creation 
and annihilation of oxygen vacancies on the surface can 
induce an insulating-to-metallic transition, which raises 
some new important questions. How large is the formation 
energy of oxygen vacancies on the \lao surface? Does 
the formation energy depend on the thickness of LaAlO$_3$? If  
thermodynamically favored, does kinetics allow the formation of 
oxygen vacancies on the surface on a reasonable time scale?  
There are likely other possible mechanisms that we have not mentioned here.

Overall, there appears to an incomplete agreement between experiment
and theory concerning the value of the critical separation. This may
suggest that the polar catastrophe at the ideal interface, while a
reasonable starting point, does not capture all the physical effects
at play, and that there can be other mechanisms that can generate
metal-insulator transitions (\eg, oxygen vacancies on the \lao
surface).

\subsection{Is there conductivity on the surface?}
\label{surface conductivity}

Based on the polar catastrophe scenario, electrons transfer from the 
surface (or capping interface) to the $n$-type interface when the 
critical separation is exceeded.  In principle, that should lead 
to two regions containing carriers and thus two conducting channels:  
electrons at the $n$-type interface and holes at the surface 
(or in the capping layers). However, to date, only electron-like 
carriers have been reported for 
the $n$-type interface. This is due in part to the fact that it is 
difficult for simple Hall measurements
to detect multi-type carriers. However, Thiel \etal
\cite{Mannhart-Science-2006} show in experiment 
that the conducting layer is 
not located on the \lao surface. 

Although further experiments are needed to confirm the insulating
behavior of the surfaces, Thiel's observation gives rise to the following 
question: if there
are holes on the surface, why do they not contribute to the transport?
One possible answer is self-trapping of the holes.  The self-trapping
can be similar in spirit to that of the self-trapped hole polaron
proposed by Pentcheva and Pickett \cite{Pickett-PRB-2006} at the
$p$-type interface. Many details remain to be clarified.
Does the presence of the surface enhance or destabilize self-trapping
tendencies?  What are the effects of surface relaxation?  Are
relatively large correlation effects (large Hubbard $U_p$) still
required for the polaron to form on the surface, and
what is the correct Hubbard $U_p$?

Cen \etal \cite{Cen-NatMat-2008} propose 
that the presence of oxygen vacancies on the surface 
can dope the interface and make it conducting. In this picture, the 
electrons associated with the oxygen vacancies on the surface 
are pushed to the interface due to the \lao polar field. There 
are no holes on the surface, and the only conducting region 
is the interface which contains the electrons.
This prediction is different from that of polar 
catastrophe mechanism, as the defects allow for a new degree of freedom.
However, experimentally, it could be challenging to 
distinguish the following two cases: whether there are holes on the 
surface that do not conduct, or whether there are no holes at all. 
Further experiments may clarify the questions 
regarding holes on the \lao surface and could simultaneously 
give indications of what mechanisms come into play for thin \lao films.

\subsection{Are there multiple types of carriers?}

The possibility of multiple types of carriers being present at the 
$n$-type interface has been suggested in both the published 
experimental \cite{Brinkman-NatMat-2007} and theoretical 
\cite{Satpathy-PRL-2008,Son-PRB-2009} literature. 
Recent experiments also invoke the same picture \cite{Seo-2009}.  
Brinkman \etal \cite{Brinkman-NatMat-2007} speculate 
that the magnetic phenomena they observe are due to spin scattering 
of itinerant electrons off of localized electrons that serve as 
localized magnetic moments.  Seo \etal \cite{Seo-2009} explain 
their optical measurements based on the simultaneous presence of 
a low density of high-mobility carriers that dominate the transport 
properties at low temperatures, together with a high density of 
low-mobility carriers localized close to the interface that dominate 
the optical spectra.  The theoretical work uses the Anderson 
localization picture in 2D to argue that the bulk of the carriers, 
which are spatially localized close to the interface and highly 2D 
in nature, do not contribute to transport due to localization.  The 
localization argument provides a possible explanation for why transport 
measurements on samples grown at higher \poo\ $\sim10^{-4}$ mbar show 
low sheet carrier densities when compared to the theoretical 
expectation of $0.5e/a_{\textrm{STO}}^2$. 

The optical approach of Seo \etal \cite{Seo-2009} is exciting, as
it has the potential to separate out different contributions from
different types of carriers. To date, these experiments have 
been performed on samples grown at low
\poo\ $\sim10^{-6}$ mbar, with correspondingly high sheet
densities of up to $3\times10^{17}$ cm$^{-2}$, beyond the
``intrinsic'' limit of $0.5e/a_{\textrm{STO}}^2$. The majority of
carriers may stem from oxygen vacancies, swamping out any signal from the
intrinsic electron gas at the interface more evident in samples grown
at higher \poo.  

Regarding the Anderson localization scenario, one experiment 
is to try to grow similar samples but 
to find a way to modify the degree of disorder at the interface.  
That could be induced by intentionally introducing a low density 
of isovalent cations during growth of the interface region or 
by introducing intentional off-stoichiometry. More quantitatively 
speaking, from the estimation in Section \ref{Anderson}, the 
required localization behavior requires mean free paths on the order of 
a few nanometers or less, which translates into point-like defects every 
nanometers as well. Since this spacing corresponds to only a few percent
areal density per interface unit cell, the localization behavior should 
show strong dependence on the density of defects introduced on the percent 
level in the interface region. On the theoretical side, 
more detailed 2D transport modeling of the $n$-type interface, 
including scattering and thus localization, can help show whether 
the picture is applicable and under what conditions of disorder.  
In addition, the expected density and scattering efficacy of 
various defects (\eg, oxygen vacancies or cation intermixtures) 
could be evaluated from first principles simulations.

\section{Conclusion}
\label{conclusion}

The SrTiO$_3$/LaAlO$_3$ interface proves to be an 
excellent example which shows the richness of new phases emerging 
at complex oxide interfaces. The dependence of each new phase on the growth 
conditions and post-annealing procedures is not yet clear and warrants 
further study, with a 
promising outlook towards various device applications of different 
functionalities. The SrTiO$_3$/LaAlO$_3$ interface paves the way 
for engineering interfaces between transition metal oxides 
and serves as an archetype of polar-nonpolar interfaces. 
With advances in thin film growth techniques, many new 
heterointerfaces will emerge \cite{Niran-PRL-2009, Wang-PRB-2009}, 
deepening our understanding of 
interface phenomena and extending the exceptional electronic 
properties in oxides.    

\begin{acknowledgments}
We acknowldege fruitful discussion or correspondence with 
Charles H. Ahn, Alexander Brinkman, Andrea Caviglia, Yoram Dagan,
Alex Demkov, Victor E. Henrich, Paul J. Kelly, Jaekwang Lee,
Jochen Mannhart, Richard M. Martin, Joe Ngai, Rossitza Pentcheva,
Warren E. Pickett, James Reiner, Nicolas Reyren, Nicola Spaldin,
Yaron Segal, Stefan Thiel, Jean-Marc Triscone, Evgeny Tsymbal,
John C. Tully, Carlos A. F. Vaz and Frederick J. Walker.
In particular, we are grateful to Charles H. Ahn, Victor E. Henrich  
and John C. Tully for their critical reading of 
our manuscript. This work is supported by the National Science
Foundation under Contract No. MRSEC DMR 0520495. The Bulldog
parallel clusters of the Yale High Performance Computing center
provide computational resources.
\end{acknowledgments}

\bibliography{sto-lao-review10}

\end{document}